\documentclass[english,reprint,unsortedaddress]{revtex4-2}
\usepackage[T1]{fontenc}
\usepackage[latin9]{inputenc}
\usepackage{mathrsfs}
\usepackage{amsmath}
\usepackage{graphicx}
\usepackage{wasysym}
\usepackage{babel}
\begin{document}
\title{Forces from stochastic density functional theory under nonorthogonal
atom-centered basis sets}
\author{Ben Shpiro}
\affiliation{Fritz Haber Center for Molecular Dynamics and Institute of Chemistry,
The Hebrew University of Jerusalem, Jerusalem 9190401, Israel}
\author{Marcel David Fabian}
\affiliation{Fritz Haber Center for Molecular Dynamics and Institute of Chemistry,
The Hebrew University of Jerusalem, Jerusalem 9190401, Israel}
\author{Eran Rabani}
\email{eran.rabani@berkeley.edu}

\affiliation{Department of Chemistry, University of California, Berkeley, California
94720, USA }
\affiliation{Materials Sciences Division, Lawrence Berkeley National Laboratory,
Berkeley, California 94720, USA }
\affiliation{The Raymond and Beverly Sackler Center of Computational Molecular
and Materials Science, Tel Aviv University, Tel Aviv 69978, Israel}
\author{Roi Baer}
\email{roi.baer@huji.ac.il}

\affiliation{Fritz Haber Center for Molecular Dynamics and Institute of Chemistry,
The Hebrew University of Jerusalem, Jerusalem 9190401, Israel}
\begin{abstract}
We develop a formalism for calculating forces on the nuclei within
the linear-scaling stochastic density functional theory (sDFT) in
a nonorthogonal atom-centered basis set representation (Fabian et
al. WIREs Comput Mol Sci. 2019;e1412. https://doi.org/10.1002/wcms.1412)
and apply it to Tryptophan Zipper 2 (Trp-zip2) peptide solvated in
water. We use an embedded-fragment approach to reduce the statistical
errors (fluctuation and systematic bias), where the entire peptide
is the main fragment and the remaining $425$ water molecules are
grouped into small fragments. We analyze the magnitude of the statistical
errors in the forces and find that the systematic bias is of the order
of $0.065\,eV/\text{Å}$ ($\sim1.2\times10^{-3}E_{h}/a_{0}$) when
$120$ stochastic orbitals are used, independently of systems size.
This magnitude of bias is sufficiently small to ensure that the bond
lengths estimated by stochastic DFT (within a Langevin molecular dynamics
simulation) will deviate by less than 1\% from those predicted by
a deterministic calculation. 
\end{abstract}
\maketitle

\section{\label{sec:Introduction}Introduction}

Kohn-Sham density functional theory (KS-DFT) is often used for estimating
the forces on the nuclei in \emph{ab-initio} molecular dynamics simulations,
with which reliable predictions concerning structure and properties
of molecules can be made. Despite the fact that it can be used to
study extended molecular systems relevant to biomolecular chemistry
and materials science,\citep{marx2009abinitio,rapaport2004theart,graziani2014frontiers,huggins2019biomolecular}
the conventional applications are limited in size due to the cubic
algorithm complexity. Therefore, several approaches to KS-DFT have
been developed and are routinely used for treating such extended systems.
These include linear-scaling approaches which rely on electron localization
within the system's interior volume,\citep{yang1991directcalculation,li1993densitymatrix,ordejon1993unconstrained,goedecker1994efficient,nunes1994generalization,wang1995ordernmultiple,hernandez1995selfconsistent,goedecker1995lowcomplexity,ordejon1996selfconsistent,bowler1997acomparison,baer1997chebyshev,palser1998canonical,goedecker1999linearscaling,scuseria1999linearscaling,galli2000largescale,adhikari2001augmented,soler2002thesiesta,skylaris2005usingonetep,gillan2007ordernfirstprinciples,ochsenfeld2007linearscaling,havu2009efficient,lin2009polebased,ozaki2010efficient,bowler2012methods,moussa2016minimax,ratcliff2017challenges,kuhne2020cp2kan,prentice2020theonetep,nakata2020largescale}
or the tight-binding DFT approach, which uses a very small basis set
complemented by approximations calibrated with empirical data,\citep{hourahine2020dftba,aradi2007dftba,elstner2001hydrogen}
and the orbital-free DFT, which is applicable to relatively homogeneous
systems.\citep{witt2018orbitalfree,karasiev2014finitetemperature}
The way many of the linear scaling approaches achieve their gentle
algorithmic complexity involves imposing a sparse structure on the
KS density matrix (DM) in a local real-space basis representation,
effectively truncating the protruding elements. The rationale of this
procedure relies on the electron localization which characterizes
many large systems.\citep{kohn1996density} However, in metallic systems
at low-temperature, and for low band semi-conductors, the localization
length is very large, and such approaches are difficult to apply.\citep{goedecker1999linearscaling}

In order to enable treatment of systems in which electron coherence
is nonlocal, a different linear scaling approach was proposed and
dubbed stochastic density functional theory (sDFT).\citep{baer2013selfaveraging}
In sDFT we use a sparse representation of the KS-DM which does not
rely on truncation or modification of its elements. Instead, sDFT
is based on the paradigm that the expectation values of the system
observables can be regarded as random variables in a stochastic process
with an expected value and a fluctuation. The fact that estimation
of electronic structure quantities can be done by statistical sampling
allows for a natural and highly effective implementation of sDFT on
parallel architectures.

The source of errors in sDFT is statistical in nature and involves
fluctuations, the magnitude of which can be controlled by statistical
sampling theory and/or by variance-reducing techniques, such as the
embedded-fragment method,\citep{neuhauser2014communication,arnon2017equilibrium,chen2019overlapped,fabian2019stochastic}
or the energy windowing approach.\citep{chen2019energywindow,chen2021stochastic}
In addition to statistical fluctuations, the sDFT estimates of the
electron density and the forces exhibit bias errors resulting from
the nonlinear nature of the SCF iterations.\citep{cytter2018stochastic,fabian2019stochastic}
The magnitude of the bias can be controlled by using the above-mentioned
variance-reducing techniques.

Early implementations of sDFT were based on real-space grid representations
of the electron density,\citep{baer2013selfaveraging,baer2013communication,arnon2017equilibrium,cytter2018stochastic,chen2019overlapped}
and were applied to relatively homogeneous systems: either to pure
bulk silicon,\citep{cytter2018stochastic,chen2019overlapped} silicon
with impurities,\citep{chen2021stochastic} H-He mixtures,\citep{cytter2019transition}
or to finite-sized hydrogen-passivated silicon nanocrystals, and water
clusters.\citep{neuhauser2016stochastic,arnon2017equilibrium,lee2020dopantlevels}
We recently demonstrated that the noisy forces produced by sDFT in
the real-space grid representation, can be used within a Langevin
dynamics approach, to determine structural properties of such large
systems. \citep{arnon2017equilibrium,arnon2020efficient} 

The real-space implementation of sDFT is especially useful as a starting
point for post-processing DFT-based methods, such as the stochastic
GW for charge excitations,\citep{neuhauser2014breaking,vlvcek2017stochastic}
the stochastic time-dependent DFT and Bethe-Salpeter equations for
neutral excitations,\citep{rabani2015timedependent,gao2015sublinear,vlvcek2019stochastic}
and for conductance calculations in warm dense matter.\citep{cytter2019transition} 

If one is only interested in the ground state atomistic structure
real-space grid representation could be quite expensive and a more
efficient representation may be beneficial. For this purpose, we recently
developed an sDFT approach based on non-orthogonal atom-centered basis
sets.\citep{fabian2019stochastic} We found that the Hamiltonian within
this compact basis has a much smaller energy range than in the real-space
grid, allowing a significant speedup of sDFT calculations. 

Despite the fact that sDFT with the non-orthogonal atom-centered
basis set is designed to address the structural properties of large
systems, up to now, we did not have the capability to estimate the
forces on nuclei and therefore focused only on the electronic energy
and density of states.\citep{fabian2019stochastic} In this paper
we develop the necessary theory and computational tools for calculating
the forces while maintaining the linear-scaling complexity of sDFT.
In addition, we analyze the statistical fluctuations and the biases
in the forces, using as a benchmark the heterogeneous system of Tryptophan
Zipper 2 (Trp-zip2) peptide solvated in water. 

The manuscript is organized as follows: In Section~\ref{sec:Formalism},
we introduce our formalism for the stochastic forces calculations.
Then, in Section~\ref{sec:Statistical-analysis}, we present the
benchmark calculations on the Tryptophan Zipper 2 (Trp-zip2) peptide
in solution. Finally, we summarize and discuss the results in Section~\ref{sec:Summary-and-conclusions}.

\section{\label{sec:Formalism}Force calculations in stochastic density functional
theory}

In this section we describe the theory of the electronic forces on
nuclei within the finite temperature KS-DFT formalism. We set the
notations and describe the basis set representation we use for Kohn-Sham
DFT in subsection~\ref{subsec:Setting-the-stage} with the combined
implementation using real space grids briefly described in subsection~\ref{subsec:Combined-real-space-grid}.
Expressions for the forces are given in subsection~\ref{subsec:Electronic-forces-on}
with a detailed derivation given in Appendix~\ref{sec:AppForces}.
Finally, in subsections~\ref{subsec:Stochastic-estimation-of}-\ref{subsec:Embedded-fragments-approach}
we provide the detail behind the stochastic evaluation of the electronic
density and any other observables in sDFT (including the forces),
and present the statistical errors involved. 

\subsection{Setting the stage \label{subsec:Setting-the-stage}}

The KS Hamiltonian is given by:
\begin{equation}
\hat{h}_{{\rm KS}}=\hat{t}_{s}+\hat{v}_{pp}^{nl}+\hat{v}_{pp}^{loc}+v_{Hxc}\left[n\right]\left(\boldsymbol{r}\right),\label{eq:Hamiltonian op}
\end{equation}
where $\hat{t}_{s}=-\frac{1}{2}\nabla^{2}$ (we use atomic units throughout
the paper) is the electron kinetic energy operator, $\hat{v}_{pp}^{nl}=\sum_{C\in nuclei}\hat{v}_{pp\left(C\right)}^{nl}$,
and $\hat{v}_{pp}^{loc}=\sum_{C\in nuclei}v_{pp\left(C\right)}^{loc}\left(\boldsymbol{\hat{r}}-\boldsymbol{R}_{C}\right)$
are the non-local and local norm-conserving pseudopotential terms
in the Kleinman-Bylander form, \citep{troullier1991efficient,kleinman1982efficacious}
for nucleus $C$, at position $\boldsymbol{R}_{C}$. The last potential
term, $\hat{v}_{Hxc}$, is the Hartree and exchange correlation potential,
depending on the electron density, $n\left(\boldsymbol{r}\right)$:
\begin{align}
v_{Hxc}\left[n\right]\left(\boldsymbol{r}\right) & =\frac{\delta\mathscr{E}_{Hxc}\left[n\right]}{\delta n\left(\boldsymbol{r}\right)}\nonumber \\
 & =\int\frac{n\left(\boldsymbol{r}'\right)}{\left|\boldsymbol{r}-\boldsymbol{r}'\right|}d^{3}r'+v_{xc}\left[n\right]\left(\hat{\boldsymbol{r}}\right),
\end{align}
 where $\mathscr{E}_{Hxc}\left[n\right]$ is the Hartree and exchange-correlation
energy functional.

We use a nonorthogonal atom-centered basis set, $\phi_{\alpha}\left(\boldsymbol{r}\right)$,
$\alpha=1,\dots,K$, with an overlap matrix $S_{\alpha\gamma}=\left\langle \phi_{\alpha}\left|\phi_{\gamma}\right.\right>$,
$\alpha,\,\gamma=1,\dots,K$.  Within such a basis set approach,
the $K\times K$ DM is given as an operator involving a function of
$HS^{-1}:$
\begin{equation}
P=S^{-1}f\left(HS^{-1};\beta,\mu\right)\label{eq:DM-as-FD}
\end{equation}
where $H_{\alpha\gamma}=\left\langle \phi_{\alpha}\left|\hat{h}_{\text{KS}}\right|\phi_{\gamma}\right\rangle $
and
\begin{equation}
f\left(\varepsilon;\beta,\mu\right)\equiv\frac{1}{1+e^{\beta\left(\varepsilon-\mu\right)}}.\label{eq:FD}
\end{equation}
is the Fermi-Dirac distribution function. The DM is used to calculate
expected values of single-electron observables $\hat{o}$ as: 
\begin{equation}
\left\langle \hat{o}\right\rangle =2\times\text{Tr}\left[OP\right],\label{eq:<o>}
\end{equation}
where $O$ is the matrix representing $\hat{o}$ in the basis, with
elements:
\begin{equation}
O_{\alpha\gamma}=\left\langle \phi_{\alpha}\left|\hat{o}\right|\phi_{\gamma}\right\rangle ,\label{eq:MatrixO}
\end{equation}
and the factor of 2 accounts for the electron's spin in a closed shell
representation. For example, the expectation value of the density
operator $\hat{n}\left(\boldsymbol{r}\right)$ is the electron density,
given by:
\begin{equation}
n\left[P\right]\left(\boldsymbol{r}\right)=\left\langle \delta\left(\boldsymbol{r}-\hat{\boldsymbol{r}}\right)\right\rangle =2\times\sum_{\alpha\gamma}P_{\alpha\gamma}\phi_{\alpha}\left(\boldsymbol{r}\right)\phi_{\gamma}\left(\boldsymbol{r}\right),\label{eq:Real-SpaceDensity}
\end{equation}
The DM in Eq.~(\ref{eq:DM-as-FD}) minimizes the total electronic
free-energy:
\begin{equation}
\Omega\left[P\right]=\mathscr{E}\left[P\right]-\mu\mathcal{N}\left[P\right]-\left(k_{B}\beta\right)^{-1}\mathcal{S}_{ent}\left[P\right].\label{eq:Free energy}
\end{equation}
Here $\mathscr{E}\left[P\right]$ is the electronic internal energy,
\begin{align*}
\text{\ensuremath{\mathscr{E}}}\left[P\right] & =2\times\text{Tr}\left[\left(T_{s}+V_{PP}^{nl}+V_{PP}^{loc}\right)P\right]\\
 & \,\,\,\,\,\,\,\,\,\,\,\,+\mathscr{E}_{Hxc}\left[n\left[P\right]\right]
\end{align*}
and the number of electrons is given by
\[
\mathcal{N}\left[P\right]=2\times\text{Tr}\left[SP\right].
\]
The actual value we use for the chemical potential $\mu$ is tuned
to enforce $\mathcal{N}\left[P\right]$ to be equal to the actual
number of electrons in the system (see Ref.\citenum{fabian2019stochastic}
for detail) . Finally $\mathcal{S}_{ent}\left[P\right]$ is the entropy
of the non-interacting electrons of the KS system, given by:
\begin{align*}
\mathcal{S}_{ent}\left[P\right] & =-2\times k_{B}\text{Tr}\left[SP\ln SP+\right.\\
 & \,\,\,\,\,\,\,\,\,\,\,\,\,\,\,\left.\left(1-SP\right)\ln\left(1-SP\right)\right]
\end{align*}
Equations (\ref{eq:Hamiltonian op})-(\ref{eq:Real-SpaceDensity})
must be solved together, and the resulting solution for the density
$n\left(\boldsymbol{r}\right)$ and the DM $P$ is called the self-consistent
field (SCF) solution to the KS equations. The procedure for reaching
SCF solution is iterative: in each iteration, called an SCF cycle,
$P$ is calculated from $H$ using Eq.~(\ref{eq:DM-as-FD}), $n\left(\boldsymbol{r}\right)$
from $P$ from which~$v_{Hxc}\left[n\right]\left(\boldsymbol{r}\right)$
is calculated and a new KS Hamiltonian matrix $H$ is built. 

\subsection{Combined real-space grid and basis set implementation\label{subsec:Combined-real-space-grid}}

The theory described in the section above uses, in addition to the
basis function $\phi_{\alpha}\left(\boldsymbol{r}\right)$, also a
Cartesian grid (with uniform grid-spacing $h$) which spans the space
occupied by the electron density. The grid is used to evaluate the
matrix elements of Eq.~(\ref{eq:MatrixO}) of various observables
$\hat{o}$, expressible as operators on the grid:
\begin{equation}
O_{\alpha\gamma}=h^{3}\sum_{g}\phi_{\alpha}\left(\boldsymbol{r}_{g}\right)\left[\hat{o}\phi_{\gamma}\right]\left(\boldsymbol{r}_{g}\right),\label{eq:gridMatrixO}
\end{equation}
where $\boldsymbol{r}_{g}$ are the grid points ($g$ is a 3D index).
Each matrix element of Eq.~(\ref{eq:gridMatrixO}) can be evaluated
efficiently\footnote{This requires a fast evaluation of basis functions $\phi_{\alpha}\left(\boldsymbol{r}_{g}\right)$
at the grid points. For this, we employ standard quantum-chemical
Cartesian functions, expressible as sums of triple products, $\phi_{\alpha}\left(x,y,z\right)=\sum_{p}\xi_{\alpha}^{p}\left(x\right)\eta_{\alpha}^{p}\left(y\right)\zeta_{\alpha}^{p}\left(z\right)$
where the sums of $\xi_{\alpha}^{p}\left(x\right)$, $\eta_{\alpha}^{p}\left(y\right)$
and $\zeta_{\alpha}^{p}\left(z\right)$ are the primitive functions
of the basis. At grid point $\boldsymbol{r}_{g}$ the basis function
is a sum (over the primitive functions) of triple products formed
from three 1D vectors: $\xi_{\alpha}^{p}\left(x_{g}\right)$, $\eta_{\alpha}^{p}\left(y_{g}\right)$
and $\zeta_{\alpha}^{p}\left(z_{g}\right)$ which are kept in memory.
The same technique is used for the evaluation of the derivatives of
the basis functions, which is relevant for the calculation of forces,
see Supporting Information, Section S1.} while we can also gain by parallel architecture, allowing different
cores to independently compute different $\alpha\gamma$ pairs. In
particular, the pseudopotentials $\hat{v}_{PP}^{nl/loc}$ are such
grid operators. Evaluating the electron density of Eq.~(\ref{eq:Real-SpaceDensity})
at the grid points allows calculation of the density-dependent Hartree
and XC potentials. For the former, we use fast Fourier transform techniques.\citep{martyna1999areciprocal} 

\subsection{Electronic forces on the nuclei \label{subsec:Electronic-forces-on}}

In this subsection we give formal expressions for the electronic forces
on the nuclei expressible as matrix trace operations, based on a finite
temperature formalism presented in Appendix~\ref{sec:AppForces}.
Our derivation and final results are similar yet differ in many ways
with those of Ref.\citenum{Niklasson2008note}. We calculate the work
done by the electrons as nucleus $C$ is displaced by $\delta_{C}X$
in the $x$-coordinate. This work is the change in the free energy
of Eq.~(\ref{eq:Free energy}), and therefore 
\begin{equation}
-F_{C}\delta_{C}X=\delta_{C}\Omega\label{eq:force def-1}
\end{equation}
where $F_{C}$, is the $x$-component of the force on the displaced
nucleus. The atom displacement $\delta_{C}X$ has three types of effects:
it causes an explicit change in its contribution to the pseudopotential
$\hat{v}_{pp}^{nl/loc}\to\hat{v}_{pp}^{nl/loc}+\delta_{C}\hat{v}_{pp}^{nl/loc}$,
it displaces the basis functions $\phi_{\alpha}\to\phi_{\alpha}+\delta_{C}\phi_{\alpha}$,
and it induces a variation in the DM, $P\to P+\delta_{C}P$, since
$P$ is required to be the minimizer of the free energy. Note that
due to this minimum principle $\delta_{C}\Omega$ is unaffected (to
first order) by $\delta_{C}P$ so that the work done on the atom (see
Appendix~\ref{sec:AppForces}), 
\begin{align}
-F_{C}\delta_{C}X & =2\times\text{Tr}\left[P\left(\delta_{C}H-\left(HS^{-1}\right)\delta_{C}S\right)\right],\label{eq: FinalExpressionWork.}
\end{align}
is given solely in terms of the variations in the Hamiltonian, 
\begin{align}
\left(\delta_{C}H\right)_{\alpha\beta} & =\left\langle \phi_{\alpha}\left|\delta_{C}\left(\hat{v}_{pp}^{nl}+\hat{v}_{pp}^{loc}\right)\right|\phi_{\beta}\right\rangle \label{eq:deltC-H}\\
 & +\left\langle \delta_{C}\phi_{\alpha}\left|\hat{h}_{KS}\right|\phi_{\beta}\right\rangle +\left\langle \phi_{\alpha}\left|\hat{h}_{KS}\right|\delta_{C}\phi_{\beta}\right\rangle \nonumber 
\end{align}
and the overlap, 
\begin{equation}
\left(\delta_{C}S\right)_{\alpha\beta}=\left\langle \delta_{C}\phi_{\alpha}\left|\phi_{\beta}\right.\right>+\left\langle \phi_{\alpha}\left|\delta_{C}\phi_{\beta}\right.\right>\label{eq:deltC-S}
\end{equation}
matrices. The first term in Eq.~(\ref{eq:deltC-H}) is the explicit
change in the pseudopotential, giving the direct forces on the atom.
The second and third terms in $\delta_{C}H$ (and similar terms in
Eq.~(\ref{eq:deltC-S}) for $\delta_{C}S$) are due to the variation
in basis functions, and they lead to the so-called Pulay forces,\citep{pulay1969abinitio}
on the atom. More details concerning the calculation of $\left(\delta_{C}S\right)_{\alpha\beta}$
and $\left(\delta_{C}H\right)_{\alpha\beta}$ are given in the Supporting
Information, Section S1. 

The estimation of the expectation value of a one-body observable $\hat{o}$,
given by Eq.~(\ref{eq:<o>}), requires the calculation of the trace
of the matrix $OP$. By definition $\text{Tr}\left[OP\right]=\sum_{k=1}^{K}\left(u^{k}\right)^{T}OPu^{k}$
where $u^{k}$ are a set of $K$ orthogonal unit vectors and the numerical
effort involves $K$ applications of $OP$ on a vector $u$, each
of which scales quadratically and thus the overall effort scales as
$O\left(K^{3}\right)$. 

One essential component in reducing the scaling of this step is to
exploit the sparsity of the $S^{-1}H$ operation on a vector $v$
\footnote{The application of $S^{-1}$ on a column vector involves repeated
applications of $S$ on the vector, within the preconditioned conjugate
gradient method, implemented in the HSL-MA61 code. HSL is a collection
of FORTRAN codes for large scale scientific computation ( http://www.hsl.rl.ac.uk/).}, which is used within a Chebyshev expansion,\citep{tal-ezer1984anaccurate}
as a Fermi-Dirac function representing $P$ (see Eq.~(\ref{eq:DM-as-FD})).
This leads to the following method for applying the DM onto a vector
$v$ \citep{goedecker1994efficient,baer1997sparsity}:
\begin{equation}
Pv=\sum_{n=0}^{N_{C}}a_{n}\left(\beta,\mu\right)v_{n},
\end{equation}
where $v_{n}$, $n=0,1,...$ is obtained recursively 
\begin{align}
v_{0} & =v\nonumber \\
v_{1} & =\left[\frac{HS^{-1}-\bar{E}}{\Delta E}\right]v_{0}\\
v_{n+1} & =2\left[\frac{HS^{-1}-\bar{E}}{\Delta E}\right]v_{n}-v_{n-1}.\nonumber 
\end{align}
Here $\frac{HS^{-1}-\bar{E}}{\Delta E}$ is a shifted-scaled operator
with eigenvalues in the interval $\left[-1,1\right]$ (so $\Delta E$
is equal to half the spectral range and $\bar{E}$ is its center).
The expansion coefficients depend on $\beta$ and $\mu$ characterizing
the Fermi-Dirac function; they rapidly decay to zero once $N_{C}$
exceeds a system size independent value determined by $\beta\Delta E$.
With this technique, the step $Pu^{k}$ involves a linear scaling
effort, and since there are $K$ such vectors the complexity of the
trace operation $\text{Tr}\left[OP\right]$ is reduced from $O\left(K^{3}\right)$
to $O\left(K^{2}\right)$.\citep{fabian2021linearscalability} 

\subsection{Stochastic estimation of observables and forces \label{subsec:Stochastic-estimation-of}}

In order to further reduce the numerical effort to linear scaling,
we use a stochastic vector approach, where the trace is sampled using
$I$ stochastic vectors instead of \emph{calculated} using a complete
set of $K$ orthonormal vectors. The calculation effort is reduced
from $O\left(K^{2}\right)$ to $O\left(IK\right)$and $I$ is system
independent. A full exposition of the method is given in Ref.~\citenum{fabian2019stochastic},
here we briefly mention the essential elements. 

Stochastic vectors $\chi^{T}=\left(\chi^{1},\dots,\chi^{K}\right)$,
have $K$ random components, $\chi^{k}$, each is a random variable
taking the values $\pm1$ with equal probability. We refer the reader
to Section S2. of the Supporting Information for definition and discussion
of random variables (collectively denoted $r$) their expected values
$\text{E}\left[r\right]$, their variance $\text{Var}\left[r\right]$,
and the statistical methods for evaluating these quantities using
finite samples. For each component of the stochastic vector: (1) $\left|\chi^{k}\right|=1$
(2) $\text{E}\left[\chi^{k}\right]=0$ and therefore $\text{Var}\left[\chi^{k}\right]=1$.
Furthermore, the product $\chi^{k}\chi^{j}$ of any pair of components
has a zero expected value ($\text{E}\left[\chi^{k}\chi^{j}\right]=0$,
$k\ne j$) and hence, in matrix form
\begin{equation}
\text{E}\left[\chi\chi^{T}\right]=\text{Id}\label{eq:stochResID}
\end{equation}
where $\text{Id}$ is the $K\times K$ identity matrix. We view Eq.~(\ref{eq:stochResID})
as the ``stochastic resolution of the identity'' and using it we
express the trace of the matrix $OP$ as $\text{Tr}\left[OP\right]=\text{Tr}\left[OP\text{E}\left[\chi\chi^{T}\right]\right]=\text{E}\left[\text{Tr}\left[OP\chi\chi^{T}\right]\right]$,
which upon rearrangement gives the stochastic trace formula:\citep{hutchinson1990astochastic}
\begin{equation}
\left\langle \hat{o}\right\rangle =2\times\text{E}\left[\chi^{T}OP\chi\right].\label{eq:Trace-Sample}
\end{equation}
The expected value $\text{E}\left[\chi^{T}OP\chi\right]$ can be estimated
using a sample of size $I$ with 
\begin{equation}
m_{I}=2\times\frac{1}{I}\sum_{i=1}^{I}\chi_{i}^{T}OP\chi_{i}\label{eq: stoch evaluation}
\end{equation}
which establishes a 70\% confidence interval $\left[m_{I}-\sigma_{I},m_{I}+\sigma_{I}\right]$
for $\left\langle \hat{o}\right\rangle $ where 
\begin{equation}
\sigma_{I}=\frac{s_{I}}{\sqrt{I}}\label{eq:uncertainty}
\end{equation}
and $s_{I}=\sqrt{\frac{1}{I-1}\sum_{i=1}^{I}\left(\chi_{i}^{T}OP\chi_{i}-m_{I}\right)^{2}}$
is the standard deviation. We would like to highlight that since Eq.~(\ref{eq: stoch evaluation})
is an average over $i$ independent $\chi_{i}^{T}OP\chi_{i}$ terms,
the computation is easily implemented to gain from parallel architecture. 

We can use the stochastic trace to estimate the electron density
at each grid point, based on Eq.~(\ref{eq:Real-SpaceDensity}). For
this, we define stochastic orbitals which are stochastic linear combinations
of the basis functions, defined on the grid as
\[
\eta_{i}\left(\boldsymbol{\boldsymbol{r}_{g}}\right)=\sum_{\alpha=1}^{K}\chi_{i}^{\alpha}\phi_{\alpha}\left(\boldsymbol{\boldsymbol{r}_{g}}\right)
\]
and \emph{projected }stochastic orbitals
\[
\xi_{i}\left(\boldsymbol{\boldsymbol{r}_{g}}\right)=\sum_{\alpha=1}^{K}\left[P\chi_{i}\right]^{\alpha}\phi_{\alpha}\left(\boldsymbol{\boldsymbol{r}_{g}}\right).
\]
Using the above we can now calculate the center of the confidence
interval for the electron density at point $\boldsymbol{r}_{g}$ as
the sample mean:
\begin{equation}
n_{I}\left(\boldsymbol{r}_{g}\right)=2\times\frac{1}{I}\sum_{i=1}^{I}\eta_{i}\left(\boldsymbol{r}_{g}\right)\xi_{i}\left(\boldsymbol{r}_{g}\right).\label{eq:nI(r)}
\end{equation}
In Ref.\citenum{fabian2019stochastic} we have presented CPU times
showing linear scaling in the calculation of sDFT observables.

The above technique can be used to evaluate the electronic forces
on the nuclei as they too are formulated as matrix traces (see Eq.~(\ref{eq: FinalExpressionWork.})).
The computational effort for evaluating the direct forces coming from
$\hat{v}_{pp}^{nl}$ (the non-local pseudopotential) as well as all
Pulay terms, for each degree of freedom, are independent of the system
size since they are local (See Supporting Information Section S1.C.
for detail). The computational effort for evaluating the force coming
from $\hat{v}_{pp}^{loc}$ (the local pseudopotential), for each degree
of freedom, will scale linearly unless specialized particle mesh methods
(beyond the scope of this paper) are used.

The SCF cycle of KS theory in sDFT involves using our best estimate
for the density, i.e. $n_{I}\left(\boldsymbol{r}\right)$ to build
the Hamiltonian. Since $n_{I}\left(\boldsymbol{r}\right)$ includes
an uncertainty (a fluctuation), the resulting Hamiltonian matrix $H$
also has a fluctuation. Then, plugging $H$ into the Chebyshev expansion
from which a new $n_{I}\left(\boldsymbol{r}\right)$ is calculated
converts the fluctuation into a bias, as discussed Section S2.C. of
the Supporting Information. Thus after the SCF converges all expectation
values have both an uncertainty $\sigma_{I}$ and a bias error, which
we define as:
\[
\Delta\rho_{I}=\left|\text{E}\left[m_{I}\right]-\left\langle \hat{o}\right\rangle ^{dDFT}\right|.
\]
The estimation of the uncertainty $\sigma_{I}$ can be done using
Eq.~(\ref{eq:uncertainty}), but the estimation of $\Delta\rho_{I}$
is more complicated since we need to determine $\text{E\ensuremath{\left[m_{I}\right]}}$.
We discuss this issue when we determine the bias error in the force
(see Section~\ref{sec:Statistical-analysis}). 

\subsection{Embedded fragments approach\label{subsec:Embedded-fragments-approach} }

In order to mitigate the fluctuation and bias errors we developed
a basis set version of the embedded-fragment (EF) approach,\citep{neuhauser2014communication,arnon2017equilibrium,chen2019overlapped,fabian2019stochastic}
which can be described in a general way as introducing a correction
term to the sDFT calculation. We first split all the atoms in the
system into $F$ fragments, such that each atom, and all basis functions
centered on it, belong to one and only one fragment. If the fragments
are chosen such that their size is independent of the total system
size, with sub-linear scaling and minimal increase in computation
time we can calculate the electron density in each fragment, using:
1.) deterministic DFT $n_{\text{dDFT}}^{f}\left(\boldsymbol{r}\right)$
($f=1,\dots,F$ ) and 2.) stochastic DFT $n_{I}^{f}\left(\boldsymbol{r}\right)$.
We then use the difference
\begin{equation}
\Delta n^{f}\left(\boldsymbol{r}\right)=n_{\text{dDFT}}^{f}\left(\boldsymbol{r}\right)-n_{I}^{f}\left(\boldsymbol{r}\right)
\end{equation}
 as a correction to the sDFT calculation of the density $n_{I}\left(\boldsymbol{r}\right)$
on the entire system:
\begin{equation}
n_{I}^{EF}\left(\boldsymbol{r}\right)=n_{I}\left(\boldsymbol{r}\right)+\sum_{f=1}^{F}\Delta n^{f}\left(\boldsymbol{r}\right).
\end{equation}
We note, that the correct result, $n_{I}^{EF}\left(\boldsymbol{r}\right)=n_{dDFT}\left(\boldsymbol{r}\right)$
is obtained in two limits: 1) when $F=1$ (i.e. the entire system
is a fragment) and 2) when $I\to\infty$, so $n_{I}^{f}\left(\boldsymbol{r}\right)\to n_{\text{dDFT}}^{f}\left(\boldsymbol{r}\right)$
etc. Similarly, the expectation value of any operator of interest,
$\hat{o}$: 
\begin{equation}
\left\langle \hat{o}\right\rangle _{I}^{EF}=\left\langle \hat{o}\right\rangle _{I}+\sum_{f}\left\langle \Delta\hat{o}^{f}\right\rangle _{I}\label{eq:frag eq}
\end{equation}
where $\left\langle \Delta\hat{o}^{f}\right\rangle _{I}=\left\langle \hat{o}^{f}\right\rangle _{\text{dDFT}}-\left\langle \hat{o}^{f}\right\rangle _{I}$.
The EF approach is applicable to the forces calculation, by choosing
$\hat{o}$ to be the relevant operators from Eq.~(\ref{eq: FinalExpressionWork.}).
For further detail on the implementation of the embedded fragments
method in our program, see Supporting Information, Section S3.

\section{\label{sec:Statistical-analysis}Statistical analysis of sDFT forces
in the Tryptophan Zipper 2 peptide}

\begin{figure*}
\centering{}\includegraphics[width=0.5\textwidth]{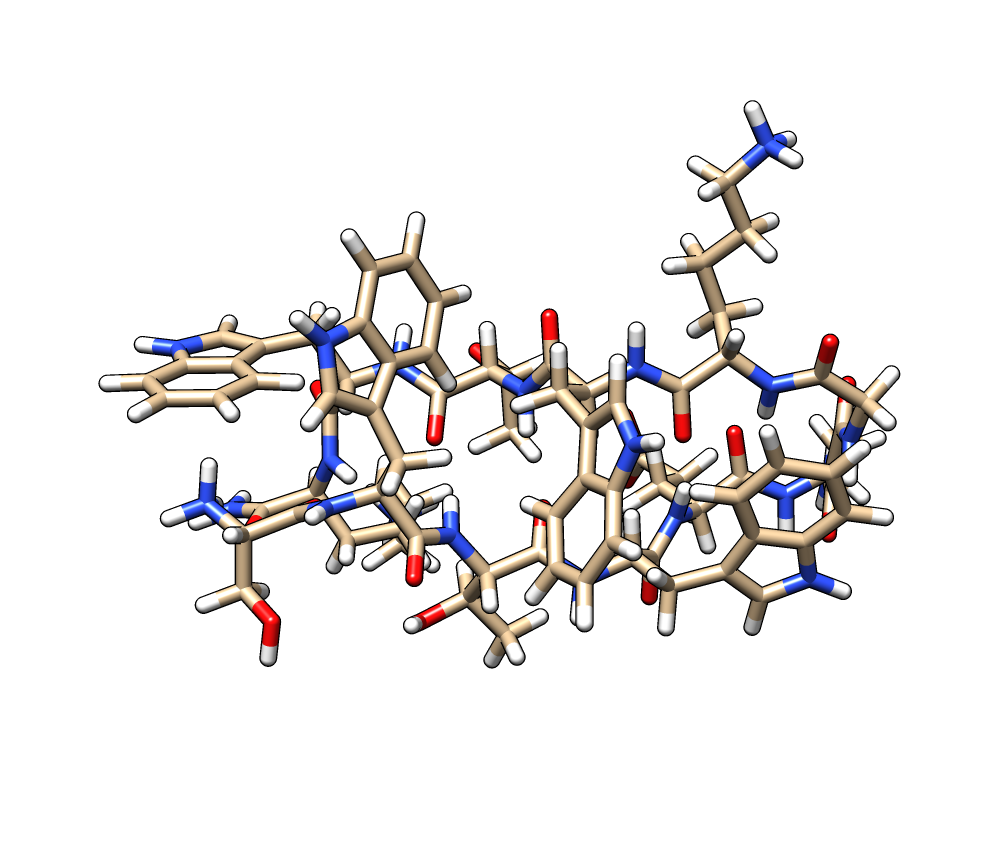}\includegraphics[width=0.5\textwidth]{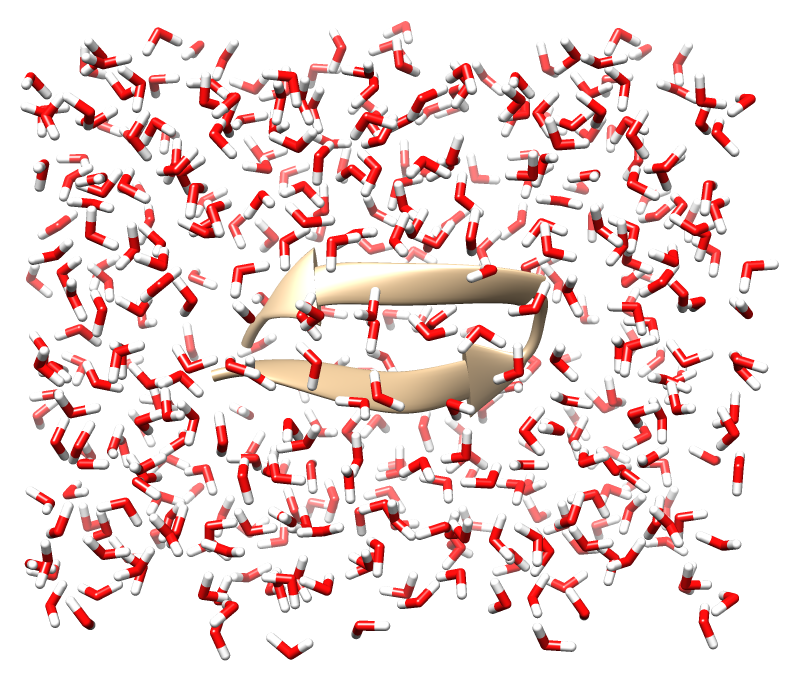}\caption{\label{fig:peptide}Left Panel: Tryptophan Zipper 2 (Trp-zip2) peptide,
composed of $220$ atoms. Right Panel: Trp-zip2 peptide (ribbon) solvated
by $425$ water molecules. The full system is composed of $1495$
atoms, $4024$ valence electrons and $3118$ basis functions are necessary
to describe it using a minimal basis set.}
\end{figure*}

Our test system is a Tryptophan Zipper 2 (Trp-zip2) peptide (pdb \emph{1le1}),
composed of $220$ atoms (left panel of Fig.~\ref{fig:peptide}),
solvated with $425$ water molecules and built using a universal force
field (UFF) in ArgusLab,\citep{thompson2004arguslab,thompson2004molecular}
(right panel of Fig.~\ref{fig:peptide}). For benchmark calculations
we focused on the $20$ nitrogen atoms of the peptide (indexed by
$C$) and calculated the forces acting on each Cartesian degree of
freedom. In these calculations, the embedded-fragment method was used,
for which we chose to consider the peptide as a single fragment and
then divided the $425$ water molecules into $27$ fragments, with
an average size of 16 molecules.

To study the statistical errors we performed the sDFT calculations
using increasing number of stochastic vectors, $I=12,120,1200$, according
to Eq.~(\ref{eq:Trace-Sample}). To estimate the magnitudes of the
bias and the uncertainty we repeated the calculations $M$ times (using
independent random number generator seeds) from which we calculated
a sample average force vector: 
\[
\bar{\boldsymbol{F}}_{C}=\frac{1}{M}\sum_{m=1}^{M}\boldsymbol{F}_{C}^{m}\:,
\]
and a $3\times3$ force covariance matrix:
\[
\Sigma^{2}=\frac{1}{M}\sum_{m=1}^{M}\left(\boldsymbol{F}_{C}^{m}-\bar{\boldsymbol{F}}_{C}\right)\left(\boldsymbol{F}_{C}^{m}-\bar{\boldsymbol{F}}_{C}\right)^{T},
\]
as an estimate for the covariance of the sDFT calculation. As the
forces acting on each atom are represented as a 3-dimensional vectors
(over the Cartesian coordinates) we would like to obtain scalar values,
irrespective of the way the Cartesian axes are defined, in order to
estimate the uncertainty and bias of the sDFT forces\footnote{In addition to the analysis given here, we also present the distribution
of the errors $F_{C}^{m}-\boldsymbol{F}_{C}^{dDFT}$ in the Supporting
Information, Section S2.D.}. For a canonical estimate of the uncertainty we use an average over
the eigenstates of $\Sigma^{2}$: 
\begin{equation}
\sigma_{C}=\sqrt{\frac{1}{3}\text{Tr}\Sigma^{2}},\label{eq:invariant sigma}
\end{equation}
where $F_{C}^{dDFT}=\left\Vert \boldsymbol{F}_{C}^{dDFT}\right\Vert $,
is the magnitude of the dDFT electronic force on atom $C$. For a
canonical estimate of the bias in the force we use the $L_{2}$-Norm
of the error in the average force vector:
\begin{align}
\Delta\rho_{C} & =\left\Vert \bar{\boldsymbol{F}}_{C}-\boldsymbol{F}_{C}^{dDFT}\right\Vert .\label{eq:invariant error}
\end{align}

\begin{figure*}[t]
\begin{centering}
\includegraphics[width=1\textwidth]{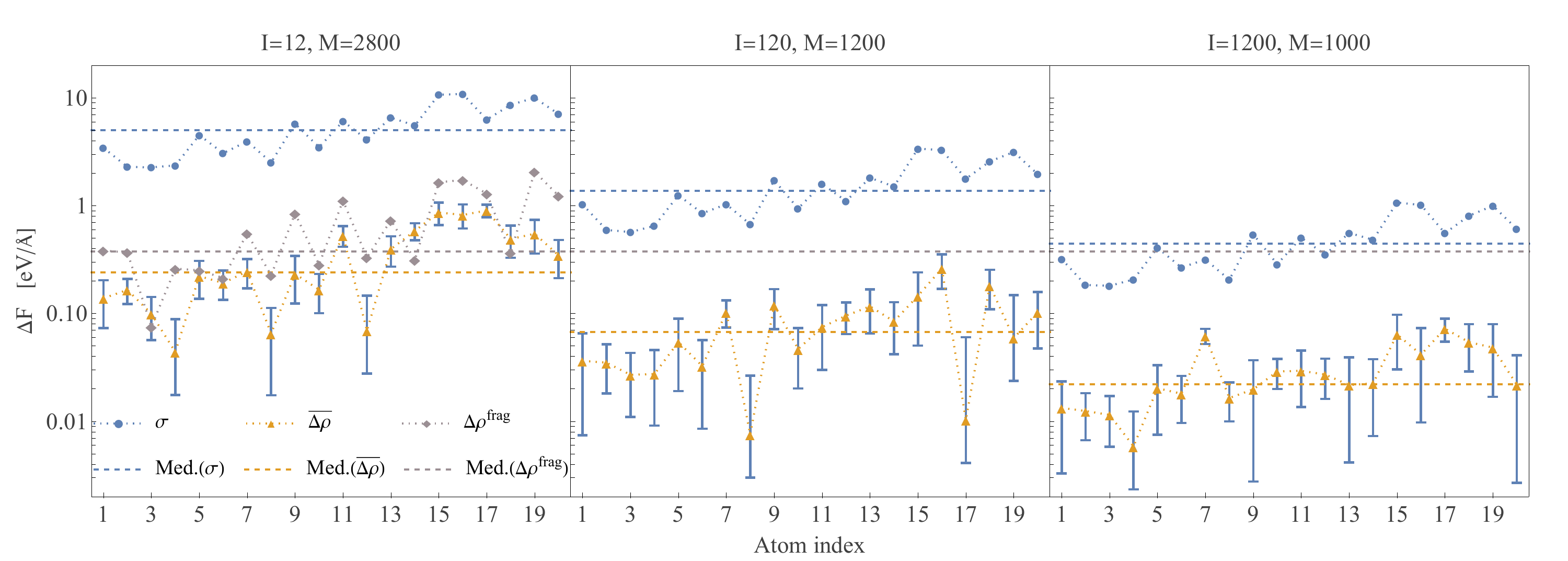}
\par\end{centering}
\caption{\label{fig:rel-force-stats}The statistical errors in the sDFT forces
acting on the $20$ nitrogen atoms in the solvated-TrpZip2 system
calculated using $I=12$,$120,1200$ stochastic vectors (see left,
center and right panels). For each Nitrogen atom, we show the uncertainty
$\sigma_{C}$ (blue dots), and the estimate in the bias $\Delta\rho_{C}$
(orange triangles), see Eqs.~\ref{eq:invariant sigma},\ref{eq:invariant error}
in text, with error bars calculated as $\pm\sigma_{C}/\sqrt{M}$.
In the $I=12$ column we also plot $\Delta\rho_{C}^{frag}=\left\Vert \boldsymbol{F}_{C}^{frag}-\boldsymbol{F}_{C}^{dDFT}\right\Vert $
(gray diamonds), where $\boldsymbol{F}_{C}^{frag}$ is the dDFT force
vector on the Nitrogen atom $C$ from the peptide-only fragment calculation.
The dotted lines connecting the markers are presented as a guide for
the eye, while the dashed horizontal lines are medians over all atoms
of $\sigma_{C}$ and $\Delta\rho_{C}$. For simplification of the
image, in the $I=120,1200$ columns we only present the median of
$\Delta\rho_{C}^{frag}$ (gray dashed line) taken over all $20$ Nitrogen
atoms.}
\end{figure*}

In Fig.~\ref{fig:rel-force-stats} we present data for the statistical
errors in the forces of the $20$ Nitrogen atoms, ordered by an atom
index according to their distance from the center of the peptide (1
closest, 20 furthest). The estimates for the uncertainty in the forces,
$\sigma_{C}$ of Eq.~(\ref{eq:invariant sigma}) are plotted in blue
circles, while the estimates of the bias $\Delta\rho_{C}$ of Eq.~(\ref{eq:invariant error}),
with an error bar calculated as $\pm\sigma_{C}/\sqrt{M}$, in orange
triangles with blue error bars. The medians over all Nitrogen atoms
are plotted as dashed lines. The used number of stochastic vectors,
$I$, as well as the number of repetitions, $M$, is shown above each
panel. We found that stable estimates of $\sigma_{C}$ are obtained
even when using a small number of $M\approx50$ repetitions and observe
that they obey the expected $1/\sqrt{I}$ behavior in accordance with
the central limit theorem. Since the variance is given by the matrix
elements of the system, (see Supporting Information, Section S2.C,
Eq.~(S3)), the pattern seen for $\sigma_{C}$ as a function of atom
index is almost unchanged for different values of $I$. To estimate
the bias we need a good estimate of $\text{E\ensuremath{\left[m_{I}\right]}}$
(the expected value of the forces when calculated using $I$ stochastic
vectors in Eq.~(\ref{eq: stoch evaluation})). As $\sigma_{C}$ is
much larger than $\Delta\rho_{C}$, a very large number of repetitions,
$M$, was required in order to achieve a good enough estimate of $\text{E\ensuremath{\left[m_{I}\right]}}$
such that $\Delta\rho_{C}$ values are useful estimates of the bias.
It is clear from the error bars that for almost all Nitrogen atoms
we have good estimates of the bias.

In the $I=12$ column, for an added perspective, we plot in gray diamonds,
the error $\Delta\rho_{C}^{frag}=\left\Vert \boldsymbol{F}_{C}^{frag}-\boldsymbol{F}_{C}^{dDFT}\right\Vert $,
where $\boldsymbol{F}_{C}^{frag}$ is the force vector on the Nitrogen
atom $C$ from a dDFT calculation on its peptide only (gas-phase)
fragment. The median is given again, in a dashed line. We observe
that the values of $\Delta\rho_{C}^{frag}$ for the atoms closer to
the center of the fragment are mostly smaller than those further away,
causing a similar pattern in the sDFT errors. When comparing the median
of $\Delta\rho_{C}^{frag}$ (plotted for all panels in a gray dashed
line) with those of the stochastic results, we see they are higher
even for the $I=12$ stochastic vectors case, whereas for the cases
of $I=120,1200$ we observe a reduction in the errors, showing that
overall sDFT significantly improves the force estimation in comparison
to the deterministic fragment calculation\footnote{We base this conclusion on the medians of $\Delta\rho_{C}$. The same
conclusions are valid also when considering the largest error, $\max_{C}\left\{ \Delta\rho_{C}\right\} $.}. 

Additional sDFT calculations on a smaller system, composed of the
Trp-zip2 peptide and only 195 solvating water molecules, show that
for a given number of stochastic orbitals ($I=12$) the uncertainty
and bias are very similar to the case of the original solvated system
(see Supporting Information, Section S4.) . This suggests the statistical
errors are roughly independent of system size. 

\section{\label{sec:Summary-and-conclusions}Summary and conclusions}

 We have presented a method for force calculations within finite
temperature sDFT in nonorthogonal atom-centered basis sets. The forces
are random variables evaluated using the stochastic trace formula
applied to various operators derived from the free energy, and are
therefore, like all sDFT observables, characterized by statistical
errors, a fluctuation and a bias. The calculation of the forces is
adapted to benefit from the embedded-fragment methodology. These calculations
are dominated by the SCF sDFT convergence step and therefore the times
for force calculations are similar to those reported in Ref.\citenum{fabian2019stochastic}. 

 In Section~\ref{sec:Statistical-analysis} we presented benchmarking
calculations, focusing on the statistical errors in the force estimates
for the $20$ Nitrogen atoms of a solvated Tryptophan Zipper 2 peptide
system. The results are given as a function of $I$, the number of
stochastic vectors used in the calculation according to Eq.~(\ref{eq:Trace-Sample}).
The uncertainty in the sDFT forces follows the expected $1/\sqrt{I}$
behavior in accordance with the central limit theorem. Using a very
large number of repetitions we were also able to uncover the bias
and determine that it is at least an order of magnitude smaller than
the uncertainty. The magnitude of the force bias is of the order of
$0.065\,eV/\text{Å}$ ($\sim10^{-3}E_{h}/a_{0}$) when $120$ stochastic
orbitals are used, independently of system size. A back-of-the-envelop
calculation shows that this magnitude of bias is sufficiently small
to ensure that the bond lengths estimated by stochastic DFT (within
a Langevin molecular dynamics simulation) will deviate by less than
1\% from those predicted by a deterministic calculation\footnote{Assuming the minimum of the Born Oppenheimer potential is harmonic
with a local force constant $k$, the bond length deviation $\delta R$
due to a force perturbation $\delta F$ obeys $\left|k\delta R\right|=\left|\delta F\right|$.
In typical solids and molecules $k$ is on the order of $5$ to $100$
$eV\text{Å}^{-2}$ \citep{gonze1998interatomic,zou2020localvibrational}
so for $\delta F$ of the order of $0.065eV/\text{Å}$ we find $\delta R\apprle0.01\,\text{Å}$,
1\% or less for most bond lengths of interest.}. Indeed, this fact was demonstrated using a Langevin Dynamics simulation
on silicon nanocrystals,\citep{arnon2017equilibrium} within a real-space
representation sDFT. Our present results indicate that sDFT based
on nonorthogonal atom-centered basis sets can be also used successfully
in this way.

 It is instructive to discuss the efficiency and accuracy of the
basis set\citep{fabian2019stochastic} vs. real-space grid\citep{baer2013selfaveraging,arnon2017equilibrium}
representations of sDFT calculations. For this, we used the $\text{Si}_{35}\text{H}_{36}$
system, comparing the 6-31G basis set calculations with those of a
real-space grid having $64^{3}$ points and grid-spacing of $\delta x=0.5a_{0}$
(for more information about this comparison see the Supporting Information,
Section S5.). We find that the time for application of the density
matrix to a random vector in the 6-31G basis is a factor 30 faster
than in the grid representation. On the other hand, surprisingly,
the standard deviation of fluctuations in a typical Si force component
is about 5 times larger in the basis set calculation than in the grid.
Therefore, we need a factor of $5^{2}=25$ more stochastic vectors
(because their number is proportional to the square of the standard
deviation) in the basis set calculation for achieving the same fluctuation
error. If we had only a single processor, the two representations
would thus require a similar numerical effort for achieving a given
fluctuation goal: the grid is 30 times slower but requires a factor
of 25 less samplings. Due to the highly parallelizable nature of sDFT,
the necessary extra sampling required by the basis-set-based calculation,
does not automatically lead to increased wall-times, if additional
CPUs can be offered. We conclude that the basis-set-based calculations
can achieve smaller wall-times than real-space grids, given additional
CPUs. 

\section*{Acknowledgments}

RB and ER gratefully thank the Binational Science Foundation grant
No. 2018368. ER acknowledges support from the Center for Computational
Study of Excited State Phenomena in Energy Materials (C2SEPEM) at
the Lawrence Berkeley National Laboratory, which is funded by the
U.S. Department of Energy, Office of Science, Basic Energy Sciences,
Materials Sciences and Engineering Division under Contract No. DE-AC02-05CH11231
as part of the Computational Materials Sciences Program.

\section*{Supporting Information Available}

Evaluation of the force matrix elements (Section S1.), basic concepts
in statistics (Section S2.), the embedded fragment method (Section
S3.), system size dependency of statistical errors (Section S4.),
efficiency of representations (Section S5.). This information is located
after the references.

\appendix

\section{Derivation of the changes in free energy \label{sec:AppForces}}

Here, we derive the force expression of Eq.~(\ref{eq: FinalExpressionWork.}).
The force is given by the change in free energy
\[
\Omega\left[P\right]=\mathscr{E}\left[P\right]-\mu\text{\ensuremath{\mathscr{N}}}\left[P\right]-\left(k_{B}\beta\right)^{-1}\mathcal{S}_{ent}\left[P\right]
\]
due to displacement of the nuclei. When nuclei are displaced the DM
also changes, we will show that under any change in the density matrix
$P\to P+\delta_{0}P$, while keeping the nuclei fixed, the free energy
of Eq.~(\ref{eq:Free energy}) does not change when $P$ is given
by Eq.~(\ref{eq:DM-as-FD}). This will be done by examining each
term in the above equation separately and summing over all of them.
Then we will consider the direct change in free energy due to a displacement
of the nuclei (while $P$ is held constant). It is only this latter
change which affects the free energy.

\subsection{The variation in $\text{\ensuremath{\mathscr{N}}}\left[P\right]$}

Starting from:
\[
\text{\ensuremath{\mathscr{N}}}\left[P\right]=\int n\left[P\right]\left(\boldsymbol{r}\right)d\boldsymbol{r}
\]
and
\[
n\left[P\right]\left(\boldsymbol{r}\right)=2\times\sum_{\alpha\gamma}P_{\alpha\gamma}\phi_{\alpha}\left(\boldsymbol{r}\right)\phi_{\gamma}\left(\boldsymbol{r}\right).
\]
Combining these we see
\[
\text{\ensuremath{\mathscr{N}}}\left[P\right]=2\times\text{Tr}\left[SP\right].
\]
We consider two types of variations: $\delta_{0}$ which change the
DM but not the atoms and $\delta_{C}$ which change the position of
atom $C$ (and thus affects the basis functions associated with that
atom) but not $P$. 
\begin{enumerate}
\item $P\to P+\delta_{0}P$ (assuming nuclei are constant): Here
\begin{equation}
\delta_{0}n\left(\boldsymbol{r}\right)=2\times\sum_{\alpha\gamma}\delta_{0}P_{\alpha\gamma}\phi_{\alpha}\left(\boldsymbol{r}\right)\phi_{\gamma}\left(\boldsymbol{r}\right)\label{eq:delta0n}
\end{equation}
so
\begin{equation}
\text{\ensuremath{\delta_{0}\mathscr{N}}}\left[P\right]=2\times\text{Tr}\left[S\delta_{0}P\right]\label{eq:delta0N}
\end{equation}
\item Nucleus C moves by $\delta_{C}X$ (and $\phi_{\alpha}\to\phi_{\alpha}+\delta_{C}\phi_{\alpha})$
(constraining $P$ to be constant): the change in the density is
\begin{align*}
\delta_{C}n\left(\boldsymbol{r}\right) & =2\times\sum_{\alpha\gamma}P_{\alpha\gamma}\left[\delta_{C}\phi_{\alpha}\left(\boldsymbol{r}\right)\phi_{\gamma}\left(\boldsymbol{r}\right)\right.\\
 & \left.+\phi_{\alpha}\left(\boldsymbol{r}\right)\delta_{C}\phi_{\gamma}\left(\boldsymbol{r}\right)\right].
\end{align*}
so:
\begin{equation}
\text{\ensuremath{\delta_{C}\mathscr{N}}}\left[P\right]=2\times\text{Tr}\left[P\delta_{C}S\right]\label{eq:deltaCN}
\end{equation}
using the change in the overlap matrix
\begin{align}
\left(\delta_{C}S\right)_{\alpha\beta} & =\left\langle \delta_{C}\phi_{\alpha}\left|\phi_{\beta}\right.\right>+\left\langle \phi_{\alpha}\left|\delta_{C}\phi_{\beta}\right.\right>\label{eq:deltaCS}
\end{align}
\end{enumerate}

\subsection{The variation in $\text{\ensuremath{\mathscr{E}}}\left[P\right]$}

Starting from:
\[
\text{\ensuremath{\mathscr{E}}}\left[P\right]=2\times\text{Tr}\left[\left(T_{s}+V_{PP}^{nl}+V_{PP}^{loc}\right)P\right]+\text{\ensuremath{\mathscr{E}_{Hxc}}}\left[n\left[P\right]\right]
\]
we have two types of variations, $\delta_{0}$ which change the DM
but not the atoms and $\delta_{C}$ which change the position of atom
$C$ (and thus affects the basis functions associated with that atom)
but not $P$.
\begin{enumerate}
\item $P\to P+\delta_{0}P$ (freezing the nuclei). We have that 
\begin{align*}
\delta_{0}\text{\ensuremath{\mathscr{E}_{Hxc}}}\left[n\left[P\right]\right] & =\int v_{Hxc}\left(n\left[P\right]\left(\boldsymbol{r}\right)\right)\delta_{0}n\left(\boldsymbol{r}\right)d\boldsymbol{r}
\end{align*}
so using Eq.~(\ref{eq:delta0n})
\begin{equation}
\delta_{0}\text{\ensuremath{\mathscr{E}}}\left[P\right]=2\times\text{Tr}\left[H\delta_{0}P\right]\label{eq:delta0E}
\end{equation}
\item Nucleus C moves by $\delta_{C}X$ (and $\phi_{\alpha}\to\phi_{\alpha}+\delta_{C}\phi_{\alpha})$
(constraining $P$ to be constant): we find
\begin{equation}
\delta_{C}\text{\ensuremath{\mathscr{E}}}\left[P\right]=2\times\text{Tr}\left[P\delta_{C}H\right]\label{eq:deltaCE}
\end{equation}
where
\end{enumerate}
\begin{align}
\left(\delta_{C}H\right)_{\alpha\beta} & =\left\langle \phi_{\alpha}\left|\delta_{C}\left(\hat{v}_{pp}^{nl}+\hat{v}_{pp}^{loc}\right)\right|\phi_{\beta}\right\rangle \label{eq:deltaCH}\\
 & +\left\langle \delta_{C}\phi_{\alpha}\left|\hat{h}_{KS}\right|\phi_{\beta}\right\rangle +\left\langle \phi_{\alpha}\left|\hat{h}_{KS}\right|\delta_{C}\phi_{\beta}\right\rangle \nonumber 
\end{align}

\subsection{The variation in $\mathcal{S}_{ent}\left[P\right]$}

Starting from $\mathcal{S}_{ent}\left[P\right]=$$-2\times k_{B}$$\text{Tr}\left[SP\ln\left(SP\right)\right.$
$\left.+\left(1-SP\right)\ln\left(1-SP\right)\right]$, 
\begin{enumerate}
\item $P\to P+\delta_{0}P$ (freezing the nuclei) We have by derivation
that 
\begin{align}
\delta_{0}\text{\ensuremath{\mathcal{S}_{ent}}}\left[n\left[P\right]\right]\label{eq:delta0Sent}\\
=-2\times k_{B} & \text{Tr}\left[\ln\left(\frac{SP}{1-SP}\right)S\delta_{0}P\right]\nonumber 
\end{align}
\item Nucleus C moves by $\delta_{C}X$ (and $\phi_{\alpha}\to\phi_{\alpha}+\delta_{C}\phi_{\alpha})$
(constraining $P$ to be constant), we find:
\end{enumerate}
\begin{align}
\delta_{C}\text{\ensuremath{\mathcal{S}_{ent}}}\left[n\left[P\right]\right] & =-2\times k_{B}\text{Tr}\left[\ln\left(\frac{SP}{1-SP}\right)P\delta_{C}S\right]\label{eq:deltaCSent}
\end{align}

\subsection{The variation in $\Omega\left[P\right]$}

Here we combine the above results, while using the relationship:
\[
\left(\mu-\beta^{-1}\ln\frac{SP}{1-SP}\right)=HS^{-1},
\]
which we find by substituting in Eq.~(\ref{eq:DM-as-FD}) for $P$. 
\begin{enumerate}
\item $P\to P+\delta_{0}P$ (freezing the nuclei) Using Eqs.~(\ref{eq:delta0N}),
(\ref{eq:delta0E}) and (\ref{eq:delta0Sent}), we have 
\begin{align*}
\delta_{0}\Omega & =2\times\\
 & \text{Tr}\left[\left\{ H-\left(\mu-\beta^{-1}\ln\frac{SP}{1-SP}\right)S\right\} \delta_{0}P\right]
\end{align*}
 leading to 
\[
\delta_{0}\Omega=0.
\]
This reflects the fact that $P$ of Eq.~(\ref{eq:DM-as-FD}) minimizes
$\Omega$$\left[P\right]$.
\item Nucleus C moves by $\delta_{C}X$ (and $\phi_{\alpha}\to\phi_{\alpha}+\delta_{C}\phi_{\alpha})$
(since a variation in $P$ does not affect the value of $\Omega$
we can take it as a constant): using Eqs.~(\ref{eq:deltaCN}), (\ref{eq:deltaCE})
and (\ref{eq:deltaCSent}), we find 
\begin{equation}
\delta_{C}\Omega=2\times\text{Tr}\left[P\left(\delta_{C}H-HS^{-1}\delta_{C}S\right)\right].\label{eq:delt-C Omega}
\end{equation}

The change in free energy is composed of two terms: a term due to
the energy, $\delta_{C}H$ (which includes a direct change and a Pulay
term, see Eq. (\ref{eq:deltaCH})), and a change due to entropy, which
depends purely on Pulay changes in the overlap matrix, $\delta_{C}S$
(see (\ref{eq:deltaCS})). 
\end{enumerate}

\bibliographystyle{aapmrev4-2}
\bibliography{forcesBib}

\end{document}

\title{Supporting Information: \\
Forces from Stochastic Density Functional Theory under Nonorthogonal
Atom-Centered Basis Sets }
\author{Ben Shpiro}
\affiliation{Fritz Haber Center for Molecular Dynamics and Institute of Chemistry,
The Hebrew University of Jerusalem, Jerusalem 9190401, Israel}
\author{Marcel David Fabian}
\affiliation{Fritz Haber Center for Molecular Dynamics and Institute of Chemistry,
The Hebrew University of Jerusalem, Jerusalem 9190401, Israel}
\author{Eran Rabani}
\email{eran.rabani@berkeley.edu}

\affiliation{Department of Chemistry, University of California, Berkeley, California
94720, United States }
\affiliation{Materials Sciences Division, Lawrence Berkeley National Laboratory,
Berkeley, California 94720, United States }
\affiliation{The Raymond and Beverly Sackler Center of Computational Molecular
and Materials Science, Tel Aviv University, Tel Aviv 69978, Israel}
\author{Roi Baer}
\email{roi.baer@huji.ac.il}

\affiliation{Fritz Haber Center for Molecular Dynamics and Institute of Chemistry,
The Hebrew University of Jerusalem, Jerusalem 9190401, Israel}
\maketitle

\section{Evaluating matrix elements for Forces \label{subsec:Evaluating-matrix-elements}}

In this section we will describe how to calculate $\left(\delta_{C}S\right)_{\alpha\beta}=\left(\frac{\partial}{\partial X_{C}}S\right)_{\alpha\beta}\delta_{C}X$:
\[
\left(\frac{\partial}{\partial X_{C}}S\right)_{\alpha\beta}=\left\langle \left.\frac{\partial}{\partial X_{C}}\phi_{\alpha}\right|\phi_{\beta}\right\rangle +\left\langle \phi_{\alpha}\left|\frac{\partial}{\partial X_{C}}\phi_{\beta}\right.\right\rangle 
\]
 and $\left(\delta_{C}H\right)_{\alpha\beta}=\left(\frac{\partial}{\partial X_{C}}H\right)_{\alpha\beta}\delta_{C}X$:
\begin{align*}
\left(\frac{\partial}{\partial X_{C}}H\right)_{\alpha\beta} & =\left\langle \phi_{\alpha}\left|\frac{\partial}{\partial X_{C}}\hat{h}_{KS}\right|\phi_{\beta}\right\rangle \\
 & +\left\langle \frac{\partial}{\partial X_{C}}\phi_{\alpha}\left|\hat{h}_{KS}\right|\phi_{\beta}\right\rangle +\left\langle \phi_{\alpha}\left|\hat{h}_{KS}\right|\frac{\partial}{\partial X_{C}}\phi_{\beta}\right\rangle 
\end{align*}
 used in Eq. 11 in the manuscript. 

\subsection{The Pulay forces \label{subsec:Pulay-terms}}

Here we give the detail of calculating the $\left\langle \left.\frac{\partial}{\partial X_{C}}\phi_{\alpha}\right|\phi_{\beta}\right\rangle $
and $\left\langle \frac{\partial}{\partial X_{C}}\phi_{\alpha}\left|\hat{h}_{KS}\right|\phi_{\beta}\right\rangle $
type matrix elements. For these we need to calculate the derivatives
of the basis functions $\phi\left(\boldsymbol{r}_{g}-\boldsymbol{R}_{C}\right)$
with respect to nuclear coordinates $R_{C}=\left(X_{C},Y_{C},Z_{C}\right)$.
Since we use Cartesian Gaussian functions as our basis we enjoy the
ease of calculating their values and derivatives analytically on the
grid. Each atom-centered basis function is a sum of primitives, $e^{-\gamma x^{2}}x^{l}\times e^{-\gamma y^{2}}y^{m}\times e^{-\gamma x^{2}}z^{n}$,
where $n+l+m$ is the total angular momentum quantum number. Therefore
each primitive can be represented by three 1-dimensional vectors $\xi_{\alpha}\left(x_{g}-X_{C}\right)$,
$\eta_{\alpha}\left(y_{g}-Y_{C}\right)$ and $\zeta_{\alpha}\left(z_{g}-Z_{C}\right)$
and the function is defined inside a ``window'' surrounding the
atom $C$ and given as a product of three terms at each of its grid
points:
\[
\phi_{\alpha}\left(\boldsymbol{r}_{g}\right)=\xi_{\alpha}\left(x_{g}-X_{C}\right)\eta_{\alpha}\left(y_{g}-Y_{C}\right)\zeta_{\alpha}\left(z_{g}-Z_{C}\right).
\]
The grid is then used to evaluate the matrix elements of the overlap
and Hamiltonian matrices:
\[
S_{\alpha\beta}=h^{3}\sum_{\boldsymbol{r}_{g}}\phi_{\alpha}\left(\boldsymbol{r}_{g}\right)\phi_{\beta}\left(\boldsymbol{r}_{g}\right),
\]
\[
H_{\alpha\beta}=h^{3}\sum_{\boldsymbol{r}_{g}}\phi_{\alpha}\left(\boldsymbol{r}_{g}\right)\left[\hat{h}_{KS}\phi_{\beta}\right]\left(\boldsymbol{r}_{g}\right),
\]
where $h$ is the uniform grid-spacing. 

Since the forces are the derivatives with respect to nuclear coordinate
$X_{C}$, while our grid points are the electronic coordinates, $x_{g}$,
we use the fact that for a basis function centered around $\boldsymbol{R}_{C}$,
taking the derivative with respect to $X_{C}$ is simply the negative
of the derivative with respect to $x$ evaluated at $\boldsymbol{r}_{g}-\boldsymbol{R}_{C}$:
\begin{align*}
\frac{\partial}{\partial X_{C}}\phi_{\alpha}\left(\boldsymbol{r}_{g}\right) & =-\xi_{\alpha}^{\prime}\left(x_{g}-X_{C}\right)\eta_{\alpha}\left(y_{g}-Y_{C}\right)\zeta_{\alpha}\left(z_{g}-Z_{C}\right),
\end{align*}
such that derivatives can therefore also be defined inside a ``window''
surrounding the atom $C$ and given as a product of three terms at
each of its grid points. The grid is then used to evaluate the Pulay
matrix elements of $\delta_{C}S$ and $\delta_{C}H$:

\[
\left\langle \left.\frac{\partial}{\partial X_{C}}\phi_{\alpha}\right|\phi_{\beta}\right\rangle =h^{3}\sum_{\boldsymbol{\boldsymbol{r}}_{g}}\left[\frac{\partial}{\partial X_{C}}\phi_{\alpha}\left(\boldsymbol{r}_{g}\right)\right]\phi_{\beta}\left(\boldsymbol{r}_{g}\right),
\]
\[
\left\langle \frac{\partial}{\partial X_{C}}\phi_{\alpha}\left|\hat{h}_{KS}\right|\phi_{\beta}\right\rangle =h^{3}\sum_{\boldsymbol{\boldsymbol{r}}_{g}}\left[\frac{\partial}{\partial X_{C}}\phi_{\alpha}\left(\boldsymbol{r}_{g}\right)\right]\left[\hat{h}_{KS}\phi_{\beta}\right]\left(\boldsymbol{r}_{g}\right),
\]
The above Pulay terms are non-zero only when the $\phi_{\alpha}$
basis belongs to atom $C$, i.e. when $\phi_{\alpha}\in C$, and therefore
need to be calculated only for a very small number of $\alpha,\beta$
pairs, i.e. only when $\phi_{\alpha},\phi_{\beta}$ have overlapping
windows on the grid. The result of these conditions is that Pulay
force matrices are incredibly sparse and their number of non-zero
elements, $N_{\text{non zero}}$, is independent of the system size,
as it depends only on the choice of basis set through the number of
basis functions per atom. As for each degree of freedom (DOF), $X_{C}$
($X$ direction of atom $C$), we have to compute a Pulay force sparse-matrix,
we exploit these conditions in our code by only evaluating non-zero
matrix elements and storing them in sparse-matrix structures. In Subsection
\ref{subsec:sparse structures} we describe the method of storage
and application (onto a vector) of the sparse structure we have used.

\paragraph{The algorithm for computing $\left(\frac{\partial}{\partial X_{C}}S\right)_{\alpha\beta}$ }
\begin{itemize}
\item For all basis functions $\alpha\in C$
\begin{itemize}
\item loop over all basis function $\beta<\alpha$ that overlap with $\alpha$,
then:
\item If $\beta\in C$
\[
\left(\frac{\partial S}{\partial X_{C}}\right)_{\alpha\beta}=\left\langle \left.\frac{\partial}{\partial X_{C}}\phi_{\alpha}\right|\phi_{\beta}\right\rangle +\left\langle \phi_{\alpha}\left|\frac{\partial}{\partial X_{C}}\phi_{\beta}\right.\right\rangle 
\]
\item otherwise
\[
\left(\frac{\partial S}{\partial X_{C}}\right)_{\alpha\beta}=\left\langle \left.\frac{\partial}{\partial X_{C}}\phi_{\alpha}\right|\phi_{\beta}\right\rangle 
\]
\end{itemize}
\item For all basis functions $\alpha\notin C$
\begin{itemize}
\item loop over all basis function $\beta<\alpha$ that overlap with $\alpha$,
then:
\item If $\beta\in C$
\[
\left(\frac{\partial S}{\partial X_{C}}\right)_{\alpha\beta}=\left\langle \phi_{\alpha}\left|\frac{\partial}{\partial X_{C}}\phi_{\beta}\right.\right\rangle 
\]
\end{itemize}
\item All other terms are not $0$, and we do not store them in the sparse
matrix structure.
\end{itemize}
(To simplify the code, as $\left\langle \phi_{\alpha}\left|\frac{\partial}{\partial X_{C}}\phi_{\beta}\right.\right\rangle =\left\langle \left.\frac{\partial}{\partial X_{C}}\phi_{\beta}\right|\phi_{\alpha}\right\rangle $,
we always take the derivative from the left side). 

Overall we get:
\[
\left(\frac{\partial S}{\partial X_{C}}\right)_{\alpha\beta}=\begin{cases}
\begin{array}{c}
\left\langle \left.\frac{\partial}{\partial X_{C}}\phi_{\alpha}\right|\phi_{\beta}\right\rangle +\left\langle \phi_{\alpha}\left|\frac{\partial}{\partial X_{C}}\phi_{\beta}\right.\right\rangle \\
\left\langle \left.\frac{\partial}{\partial X_{C}}\phi_{\alpha}\right|\phi_{\beta}\right\rangle \\
\left\langle \phi_{\alpha}\left|\frac{\partial}{\partial X_{C}}\phi_{\beta}\right.\right\rangle \\
0
\end{array} & \begin{array}{c}
\alpha,\beta\in C\\
\alpha\in C,\beta\notin C\\
\alpha\notin C,\beta\in C\\
\alpha,\beta\notin C
\end{array}\end{cases}
\]
and the same can be done for the Hamiltonian Pulay terms:

\[
\left(\frac{\partial H}{\partial X_{C}}\right)_{\alpha\beta}^{\text{Pulay}}=\begin{cases}
\begin{array}{c}
\left\langle \frac{\partial}{\partial X_{C}}\phi_{\alpha}\left|\hat{h}_{KS}\right|\phi_{\beta}\right\rangle +\left\langle \phi_{\alpha}\left|\hat{h}_{KS}\right|\frac{\partial}{\partial X_{C}}\phi_{\beta}\right\rangle \\
\left\langle \frac{\partial}{\partial X_{C}}\phi_{\alpha}\left|\hat{h}_{KS}\right|\phi_{\beta}\right\rangle \\
\left\langle \phi_{\alpha}\left|\hat{h}_{KS}\right|\frac{\partial}{\partial X_{C}}\phi_{\beta}\right\rangle \\
0
\end{array} & \begin{array}{c}
\alpha,\beta\in C\\
\alpha\in C,\beta\notin C\\
\alpha\notin C,\beta\in C\\
\alpha,\beta\notin C
\end{array}\end{cases}
\]
however, as the Hamiltonian includes terms that are explicitly dependent
on nuclear coordinates in the form of the non-local (\emph{nl}) and
local (\emph{loc}) pseudopotential terms: $\hat{v}_{pp}^{nl/loc}=\sum_{C'\in nuclei}\hat{v}_{pp\left(C'\right)}^{nl/loc}$,
there are also $\left\langle \phi_{\alpha}\left|\frac{\partial}{\partial X_{C}}\hat{h}_{KS}\right|\phi_{\beta}\right\rangle =\left\langle \phi_{\alpha}\left|\frac{\partial}{\partial X_{C}}\hat{v}_{pp}^{nl/loc}\right|\phi_{\beta}\right\rangle $
type terms that contribute to the overall force on atom $C$ in the
$X$ direction$.$ See Subsection \ref{subsec:direct force (nl and loc)}
below for the detail of these force terms.

\subsection{The direct forces \label{subsec:direct force (nl and loc)}}

Here we give the detail of calculating the $\left\langle \phi_{\alpha}\left|\frac{\partial}{\partial X_{C}}\hat{v}_{pp}^{nl/loc}\right|\phi_{\beta}\right\rangle $
type matrix elements. 

Since the non-local and local pseudopotential operators are $\hat{v}_{pp}^{nl/loc}=\sum_{C'\in nuclei}\hat{v}_{pp\left(C'\right)}^{nl/loc}$,
the derivative with respect to a variation in nuclear coordinate $X_{C}$is
given by: 
\[
\frac{\partial}{\partial X_{C}}\hat{v}_{pp}^{nl/loc}\left|\phi_{\beta}\right\rangle =\frac{\partial}{\partial X_{C}}\hat{v}_{pp\left(C\right)}^{nl/loc}\left|\phi_{\beta}\right\rangle .
\]
The $\hat{v}_{pp\left(C\right)}^{nl/loc}$ operators have an analytical
expression of the Kleinman-Bylander form \citep{kleinman1982efficacious},
such that we can apply it, and its derivative, $\frac{\partial}{\partial X_{C}}\hat{v}_{pp\left(C\right)}^{nl/loc}$,
on a vector on the grid. 

Due to its short-range nature, $\hat{v}_{pp\left(C\right)}^{nl}$
and subsequently $\frac{\partial}{\partial X_{C}}\hat{v}_{pp\left(C\right)}^{nl}$
are stored on a small ``window'' of grid points around $R_{C}$.
The $\left\langle \phi_{\alpha}\left|\frac{\partial}{\partial X_{C}}\hat{v}_{pp\left(C\right)}^{nl}\right|\phi_{\beta}\right\rangle $
matrix elements are therefore calculated as a multiplication of two
grid vectors: 
\begin{equation}
\left(\frac{\partial}{\partial X_{C}}V_{pp\left(C\right)}^{nl}\right)_{\alpha\beta}^{\text{direct}}=h^{3}\sum_{\boldsymbol{r}_{g}\in\left(\alpha\cap\beta\cap\hat{v}_{pp\left(C\right)}^{nl}\right)}\phi_{\alpha}\left(\boldsymbol{r_{g}}\right)\left[\left[\frac{\partial}{\partial X_{C}}\hat{v}_{pp\left(C\right)}^{nl}\right]\phi_{\beta}\right]\left(\boldsymbol{r_{g}}\right),\label{eq:deltaVnl matrix elements}
\end{equation}
where the sum over grid points $\boldsymbol{r}_{g}$ is over only
grid points that are inside the windows of all three terms, $\phi_{\alpha}$,
$\phi_{\beta}$and $\hat{v}_{pp\left(C\right)}^{nl}$. This requirement
of overlapping windows for all three terms results in very sparse
matrices, with the number of non-zero matrix elements, $N_{\text{non zeros}}$,
dependent only on the choice of basis set and independent of system
size.

For each degree of freedom (DOF), $X_{C}$ (atom $C$ in the $X$
direction), we have to compute a $\frac{\partial}{\partial X_{C}}V_{pp\left(C\right)}^{nl}$
force sparse-matrix and apply it to a stochastic vector as part of
the evaluation using the stochastic trace formula. In Subsection \ref{subsec:sparse structures}
we describe the method of storage and application of the sparse structure
we have used.

\paragraph{The algorithm for computing $\left(\frac{\partial}{\partial X_{C}}V_{pp\left(C\right)}^{nl}\right)_{\alpha\beta}^{\text{direct}}$ }
\begin{itemize}
\item loop over all basis functions $\beta$ that have overlapping windows
with $\hat{v}_{pp\left(C\right)}^{nl}$ 
\begin{itemize}
\item Calculate the $\left[\frac{\partial}{\partial X_{C}}\hat{v}_{pp\left(C\right)}^{nl}\phi_{\beta}\right]\left(\boldsymbol{r}_{g}\right)$
grid vector for all $\boldsymbol{r}_{g}\in\beta\cap\hat{v}_{pp\left(C\right)}^{nl}$ 
\begin{itemize}
\item loop over all basis functions $\alpha$ that overlap with both $\beta$,
$\hat{v}_{pp\left(C\right)}^{nl}$
\begin{itemize}
\item Calculate $\left(\frac{\partial}{\partial X_{C}}V_{pp\left(C\right)}^{nl}\right)_{\alpha\beta}$
according to Eq.(\ref{eq:deltaVnl matrix elements})
\end{itemize}
\end{itemize}
\end{itemize}
\end{itemize}
The local PP operator, $\hat{v}_{pp}^{loc}$, (which includes the
long range Coulomb attraction), depends directly on the density $n\left(\boldsymbol{r}_{g}\right)$.
In this case, an equivalent method to the trace calculation of the
direct $\frac{\partial}{\partial X_{C}}\hat{v}_{pp}^{loc}$ is done
by usual planewaves calculations on the grid using $n\left(\boldsymbol{r}_{g}\right)$
in reciprocal space. This scaling of this approach is quadratic with
system size when the forces on all atoms are calculated, however it
is highly efficient.

\subsection{Sparse matrix structures for Pulay and non-local PP forces\label{subsec:sparse structures}}

To exploit the sparsity of the above, direct and Pulay, force matrices,
we do not store them in full $K\times K$ matrix structures (where
$K$ is the number of basis functions), but rather use compact a storage
structure. In these, for every degree of freedom we store three lists
of length $N_{\text{non zeros}}$: $\left(\text{val,I,J}\right)\equiv M_{ij}$
where $\left(\text{I,J}\right)$ gives the location of the value,
$\text{val}$, in the $K\times K$ matrix $M$.
\[
M=\left(\begin{array}{cccc}
0 & \cdots & 0 & 0\\
\vdots & \ddots & \vdots & \vdots\\
0 & \cdots & a & b\\
0 & \cdots & c & d
\end{array}\right)_{K\times K}\equiv\stackrel{\text{val}}{\left(\begin{array}{c}
a\\
b\\
c\\
d
\end{array}\right)}\stackrel{\text{I}}{\left(\begin{array}{c}
K-1\\
K-1\\
K\\
K
\end{array}\right)}\stackrel{\text{J}}{\left(\begin{array}{c}
K-1\\
K\\
K-1\\
K
\end{array}\right)}
\]
As per the above example, the sparse structure allows for a significant
reduction in the number of stored values in memory as we only store
$\left(3\times N_{\text{non zeros }}\right)$ elements per DOF as
opposed to $K^{2}$ elements per DOF. Since the number of DOF's, $N_{\text{DOF }}\propto K$
and since $N_{\text{non zeros }}$is a small number dependent only
on the choice of basis set (but not on system size!) our sparse structure
reduces the scaling of the memory requirement from $\mathcal{O}\left(K^{3}\right)$
to $\mathcal{O}\left(K\right).$

The sparse structure also allows for an efficient matrix vector multiplication.
For a matrix $M$ and vector $\left|\boldsymbol{z}\right\rangle $
the operation 
\[
\left|\boldsymbol{y}\right\rangle =M\left|\boldsymbol{z}\right\rangle 
\]
is given by:
\[
y_{k}=\sum_{n,\,I\left(n\right)=k}\text{val}\left(n\right)\times z_{\text{J}\left(n\right)}
\]
such that we only require $N_{\text{non zeros }}$multiplications
for a matrix vector operation (as all terms of $c$, that are not
in the $I$ list, are zero).

For the stochastic trace formula we need to calculate expectation
values using the stochastic vectors, $\chi$. In a bra-ket notation,
for a matrix $M$:
\[
r=\left\langle \chi\left|M\right|\chi\right\rangle 
\]
is given by:
\[
r=\sum_{n}^{N_{\text{non zeros}}}\chi_{\text{I}\left(n\right)}\times\text{val}\left(n\right)\times\chi_{\text{J}\left(n\right)}
\]
such that we only require $2\times N_{\text{non zeros }}$multiplications.
Since the calculation of the force operators' expectation values (direct
and Pulay), per DOF, are independent of system size, the overall scaling
of the force calculations using this sparse structure is $\mathcal{O}\left(K\right)$. 

\section{Basic concepts in statistics}

\subsection{\label{subsec:Random-variables}Random variables }

In order to understand the statistical errors involved in our procedures
we briefly review the concept of a random variable $r$ \citep{papoulis2002probability}.
It takes any one of a discrete set of values $\left\{ r\right\} $
with a given probability $p\left(r\right)\ge0$, where $\sum_{r}p\left(r\right)=1$.
The expected value of $r$ is: $\text{E}\left[r\right]=\sum_{r}rp\left(r\right)$
and the variance is $\text{Var}\left[r\right]=\text{E}\left[\left(r-\text{E}\left[r\right]\right)^{2}\right]$.
Using a sample of $I$ independent draws from the population of $r$'s
we calculate the mean 
\[
m_{I}=\frac{1}{I}\sum_{i=1}^{I}r_{i}
\]
 and the standard deviation 
\[
s_{I}=\sqrt{\frac{1}{I-1}\sum_{i=1}^{I}\left(r_{i}-m_{I}\right)^{2}},
\]
both can also be viewed as random variables with appropriate probability
functions themselves. It can be shown that
\begin{enumerate}
\item The expected value of $m_{I}$ is the same as that of $r$: 
\[
\text{E}\left[m_{I}\right]=\text{E\ensuremath{\left[r\right]}}
\]
\item The variance of $m_{I}$ is smaller by a factor $I$ than that of
$r$: 
\[
\text{Var}\left[m_{I}\right]=\frac{1}{I}\text{Var}\left[r\right]
\]
\item The expected value of $s_{I}^{2}$ is equal to the variance of: 
\[
\text{E}\left[s_{I}^{2}\right]=\text{Var}\left[r\right].
\]
\end{enumerate}
From these properties, $m_{I}$ and $s_{I}^{2}$ can serve as unbiased
estimators of the expected value and variance of the original random
variable. When $I$ is sufficiently large, the interval of values
$\left[m_{I}-\sigma_{I},m_{I}+\sigma_{I}\right]$, where 
\[
\sigma_{I}=\frac{s_{I}}{\sqrt{I}}
\]
is the uncertainty giving a 70\% confidence interval for $\text{E}\left[r\right]$.
Based on the sampled data, there is a probability of 70\% that $E\left[r\right]$
falls within this interval. 

\subsection{Stochastic vectors}

In section 2.4 of the main text we discuss the stochastic evaluation
of observables using the stochastic trace formula

\begin{equation}
\text{Tr}\left[A\right]=\text{E}\left[\chi^{T}A\chi\right],\label{eq:StochTrace}
\end{equation}
where we treat each $\chi^{T}A\chi$ as a random variable. The variance
associated the result is given by
\begin{equation}
\text{Var}\left[\chi^{T}A\chi\right]=\frac{1}{2}\sum_{i\ne j}\left(A_{ij}+A_{ji}\right)^{2}.\label{eq:variance x^T A X}
\end{equation}
The relation in Eq.~(\ref{eq:StochTrace}) is called the stochastic
trace formula and it allows evaluating the trace of $A$ by parameter
estimation techniques based on statistical sampling theory. 

\paragraph{Proof of Eq. (\ref{eq:variance x^T A X})}

We begin the proof using the definition of the variance of a random
variable, $\text{Var}\left[x\right]=\text{E}\left[x^{2}\right]-\text{E}\left[x\right]^{2}$,
and considering $\chi^{T}A\chi$ as our random variable: 
\begin{align}
\text{Var}\left[\chi^{T}A\chi\right] & =\text{E}\left[\left(\chi^{T}A\chi\right)^{T}\chi^{T}A\chi\right]-\left(\text{E}\left[\chi^{T}A\chi\right]\right)^{2}\nonumber \\
 & =\text{E}\left[\chi^{T}A^{T}\chi\chi^{T}A\chi\right]-\left(\text{Tr}\left[A\right]\right)^{2}\nonumber \\
 & =\text{E}\left[\chi_{k}\chi_{l}\chi_{i}\chi_{j}\right]A_{kl}A_{ij}-\left(\text{Tr}\left[A\right]\right)^{2}\label{eq:app-varA}
\end{align}
where in the second line we have used Eq.~(\ref{eq:StochTrace}),
and in the third line the fact that the matrix $A$ is completely
deterministic. We will now evaluate $\text{E}\left[\chi_{k}\chi_{l}\chi_{i}\chi_{j}\right]$,
using that $\text{E}\left[\chi_{i}\chi_{j}\right]=\delta_{ij}$:
\begin{align*}
\text{E}\left[\chi_{k}\chi_{l}\chi_{i}\chi_{j}\right] & =\delta_{kl}\text{E}\left[\chi_{i}\chi_{j}\right]+\left(1-\delta_{kl}\right)\left(\delta_{ki}\text{E}\left[\chi_{l}\chi_{j}\right]+\left(1-\delta_{ki}\right)\left(\delta_{kj}\text{E}\left[\chi_{i}\chi_{l}\right]\right)\right)\\
 & =\delta_{kl}\delta_{ij}+\left(1-\delta_{kl}\right)\left(\delta_{ki}\delta_{lj}+\left(1-\delta_{ki}\right)\delta_{kj}\delta_{il}\right)\\
 & =\delta_{kl}\delta_{ij}+\delta_{ki}\delta_{lj}+\delta_{kj}\delta_{il}-2\delta_{kl}\delta_{kj}\delta_{ik}
\end{align*}
and multiply by $A_{kl}A_{ij}$ and sum over all indices:
\begin{align*}
\text{E}\left[\chi_{k}\chi_{l}\chi_{i}\chi_{j}\right]A_{kl}A_{ij} & =A_{kk}A_{ii}+A_{ij}A_{ij}+A_{ji}A_{ij}-2A_{ii}^{2}\\
 & =\text{Tr}\left[A\right]^{2}+\sum_{i\ne j}A_{ij}\left(A_{ij}+A_{ji}\right)\\
 & =\text{Tr}\left[A\right]^{2}+\frac{1}{2}\sum_{i\ne j}\left(A_{ij}+A_{ji}\right)^{2}
\end{align*}

Substituting back into Eq.~(\ref{eq:app-varA}), we arrive at 
\begin{align*}
\text{Var}\left[\chi^{T}A\chi\right] & =\frac{1}{2}\sum_{k\ne l}\left(A_{kl}+A_{lk}\right)^{2}
\end{align*}

\subsection{Parameter estimation and statistical errors}

Often expected value $\text{E}\left[r\right]$ of a distribution of
a random variable $r$ is not known. The estimation of this parameter
can be done, based on the use of a finite sample $r_{i}$ of size
$I$, as discussed in section \ref{subsec:Random-variables}. As an
demonstration of this procedure we return to the question of how to
evaluate the $\text{Tr}\left[A\right]$. We take a sample of $I$
stochastic vectors $\chi_{i}$ and form a random variable $\frac{1}{I}\sum_{i=1}^{I}\chi_{i}^{T}A\chi_{i}$.
Then 
\begin{equation}
\text{Tr}\left[A\right]=\text{E}\left[\frac{1}{I}\sum_{i=1}^{I}\chi_{i}^{T}A\chi_{i}\right],\label{eq:StochTrace-1}
\end{equation}
and
\[
\text{Var}\left[\frac{1}{I}\sum_{i=1}^{I}\chi_{i}^{T}A\chi_{i}\right]=\frac{1}{2I}\sum_{k\ne l}\left(A_{kl}+A_{lk}\right)^{2}.
\]
As discussed above, the sample mean $m_{I}=\frac{1}{I}\sum_{i=1}^{I}\chi_{i}^{T}A\chi_{i}$
and corresponding standard deviation $s_{I}$ can be used to provide
a confidence interval of uncertainty $\sigma_{I}=s_{I}/\sqrt{I}$
for the value of $\text{Tr}\left[A\right]$. This statistical approach,
of building a confidence interval for $\text{Tr}\left[A\right]$ involves
$I$ applications of $A$ to a vector, whereas the deterministic calculation
of $\text{Tr}\left[A\right]$ involves $K$ such applications. Therefore,
as long as $I\ll K$ we obtain a large saving in the numerical effort,
but at the price of introducing an uncertainty. 

Now, suppose we wanted to estimate a given function of the expected
value of a random variable, $f\left(\text{E}\left[r\right]\right)$
. The simplest procedure is apply $f$ to the sample mean $m_{I}$
and take $f\left(m_{I}\right)$ as such an estimate. This procedure
works when $f\left(x\right)$ is a linear function of $x$ but otherwise
will generally incur a systematic error, called a bias. For example,
when $f\left(x\right)=x^{2}$ and $r$ is a random variable with $\text{E}\left[r\right]=0$
and $\text{Var}\left[r\right]=1$, then $\text{E}\left[f\left(m_{I}\right)\right]=\text{E}\left[m_{I}^{2}\right]=\frac{\text{E}\left[r^{2}\right]}{I}=\frac{1}{I}$
which is clearly different from the exact value of $f\left[\text{E}\left[r\right]\right]=\text{E}\left[r\right]^{2}=0$.
Hence we have the undesirable case, that for a finite value of $I$,
errors will involve fluctuations around the wrong value. Note however,
that as $I$ grows, the bias diminishes in proportion to $\frac{1}{I}$.
To be useful, when a bias exists, we need to make sure it is of sufficiently
small magnitude. 

\subsection{Distribution of random errors in the forces}

\begin{figure}
\centering{}\includegraphics[width=1\textwidth]{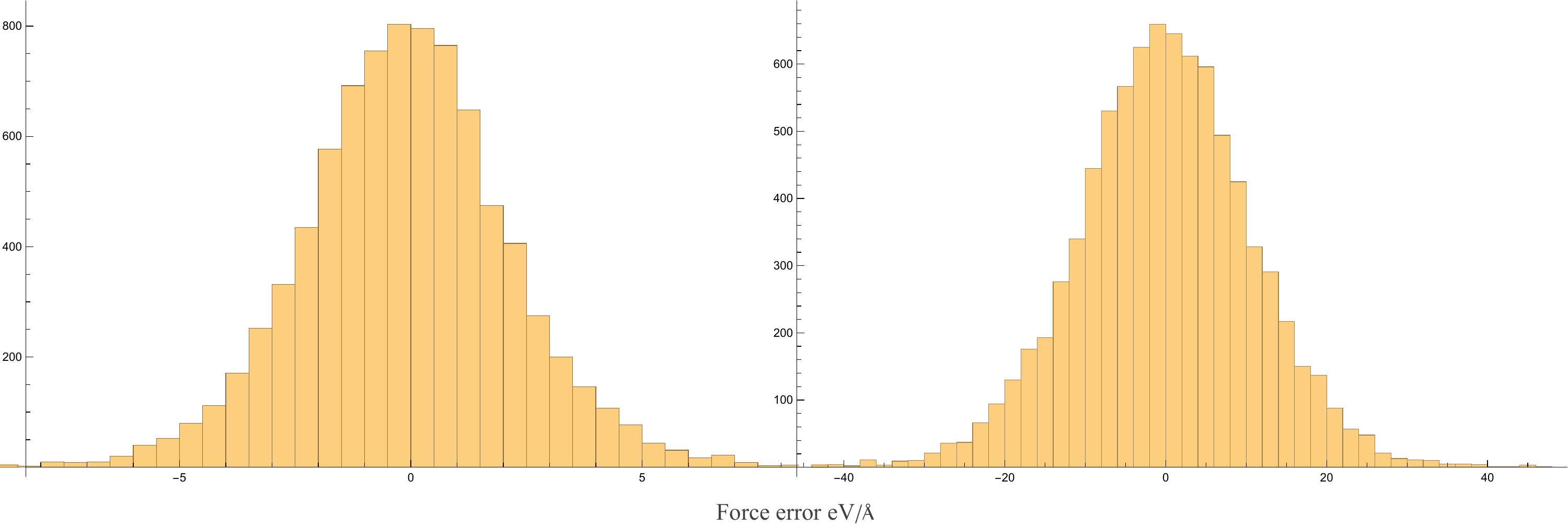}\caption{\textcolor{blue}{\label{fig:error distribution} }Histograms representing
the distribution in the errors in the sDFT forces with $I=12$ stochastic
vectors. We show the errors of forces acting on two nitrogen atoms
from the solvated-TrpZip2 system. On the left (right) panel is the
data for the smallest (largest) standard deviation cases from all
nitrogen atoms. The data for the histogram was collected by repeating
the sDFT force calculation $M=2800$ times and comparing the force
vector $\boldsymbol{F}_{C}^{m}$ on each atom $C$ to the corresponding
deterministic (accurate) value $\boldsymbol{F_{C}^{dDFT}}$. Thus
we obtained $3\times M$ error values in the x-y-z components of the
force, which represent the scatter of force components for the atom.}
\end{figure}

In Fig. \ref{fig:error distribution} we present histograms of the
sDFT force errors ($\boldsymbol{F}_{C}^{m}-\boldsymbol{F}_{C}^{dDFT},\,\,m=1,\dots,M)$
on two nitrogen atoms in the solvated-TrpZip2 system. We selected
the atoms that have the smallest/largest standard deviation. For both
atoms, we find a Gaussian-looking distribution of the force errors
centered around zero with a standard deviation within the bounds reported
in the Manuscript. 

\section{Embedded fragments\label{sec:Embedded-fragments}}

\subsection{Calculation detail in the sDFT code}

We would like to explain here how the equation: 
\begin{equation}
\left\langle \hat{o}\right\rangle _{I}^{EF}=\left\langle \hat{o}\right\rangle _{I}+\sum_{f}\left\langle \Delta\hat{o}^{f}\right\rangle _{I}\label{eq:frag eq}
\end{equation}
is implemented within our sDFT code. We give here the steps in the
calculation of the correction terms from each fragment,$\left\langle \Delta\hat{o}^{f}\right\rangle _{I}$,
where we defined the fragment-based correction as the difference between
the dDFT and sDFT, for every fragment:
\[
\left\langle \Delta\hat{o}^{f}\right\rangle _{I}=\left\langle \hat{o}^{f}\right\rangle _{dDFT}-\left\langle \hat{o}^{f}\right\rangle _{I}
\]
For the calculation of $\left\langle \hat{O}\right\rangle _{\text{dDFT}}^{f}$
we solve the generalised eigenvalue problem, find $P^{f}$, the density
matrix of the fragment subsystem, and trace:
\begin{equation}
\left\langle \hat{o}\right\rangle _{\text{dDFT}}^{f}=2\:\text{Tr}\left[\left(OP\right)^{f}\right]\label{eq:det f trace}
\end{equation}
while $\left\langle \hat{o}\right\rangle _{I}^{f}$ is calculated
using the stochastic trace formula:
\[
\left\langle \hat{o}\right\rangle _{I}^{f}=2\:\frac{1}{I}\sum_{i=1}^{I}\left(\chi_{i}^{T}\right)^{f}\left(OP\right)^{f}\chi_{i}^{f}
\]
with the vectors $\chi_{i}^{f}$ that are ``cut-outs'' of the vectors
$\chi_{i}$ which we use in the stochastic trace formula for the full
system, such that the stochastic element in the fragment calculation
corresponds with the one made on the full system. Rearranging the
above equation to be written as a trace expression gives:
\[
\left\langle \hat{o}\right\rangle _{I}^{f}=2\:\text{Tr}\left[\left(OP\right)^{f}\left(\frac{1}{I}\sum_{i=1}^{I}\chi_{i}^{f}\left(\chi_{i}^{T}\right)^{f}\right)\right]
\]
Finally, for the deterministic trace of Eq.~(\ref{eq:det f trace})
we multiply $\left(OP\right)^{f}$ by $\left(\text{Id}\right)^{f}$,
the identity matrix of the fragment dimensions, which allows us to
rewrite the fragment correction as:
\begin{align*}
\left\langle \Delta\hat{o}^{f}\right\rangle _{I} & =2\:\text{Tr}\left[\left(OP\Delta_{I}\right)^{f}\right]
\end{align*}
where $\Delta_{I}^{f}=\left(\text{Id}\right)^{f}-\frac{1}{I}\sum_{i=1}^{I}\chi_{i}^{f}\left(\chi_{i}^{T}\right)^{f}$
. 

\subsection{Another look at the fragments}

A second useful outlook on the embedded-fragments method gives insight
into the reason the statistical errors are reduced when the method
is used. We rearrange Eq. (\ref{eq:frag eq}) to get: 
\begin{align}
\left\langle \hat{o}\right\rangle _{I}^{EF}=\sum_{f}\left\langle \hat{o}\right\rangle _{\text{dDFT}}^{f}+ & \left[\left\langle \hat{o}\right\rangle _{I}-\sum_{f}\left\langle \hat{o}\right\rangle _{I}^{f}\right]\label{eq: stoch correction}\\
=2\:\text{Tr}^{det}\left[\sum_{f}\left(OP\right)^{f}\right] & +2\:\text{Tr}^{stoch}\left[O\left(P-\sum_{f}P^{f}\right)\right]\nonumber 
\end{align}
such that$\text{Tr}^{stoch}\left[O\left(P-\sum_{f}P^{f}\right)\right]$
is a stochastic correction to the dDFT calculation summed over all
fragments. As per Eq.~(\ref{eq:variance x^T A X}) the variance in
the stochastic trace is given by the magnitude of the off-diagonal
matrix elements of $O\left(P-\sum_{f}P^{f}\right)$, and so clearly,
as $\sum_{f}P^{f}$ approaches $P$ the stochastic trace will have
much smaller variance.

\begin{figure}
\centering{}\includegraphics[width=0.9\textwidth]{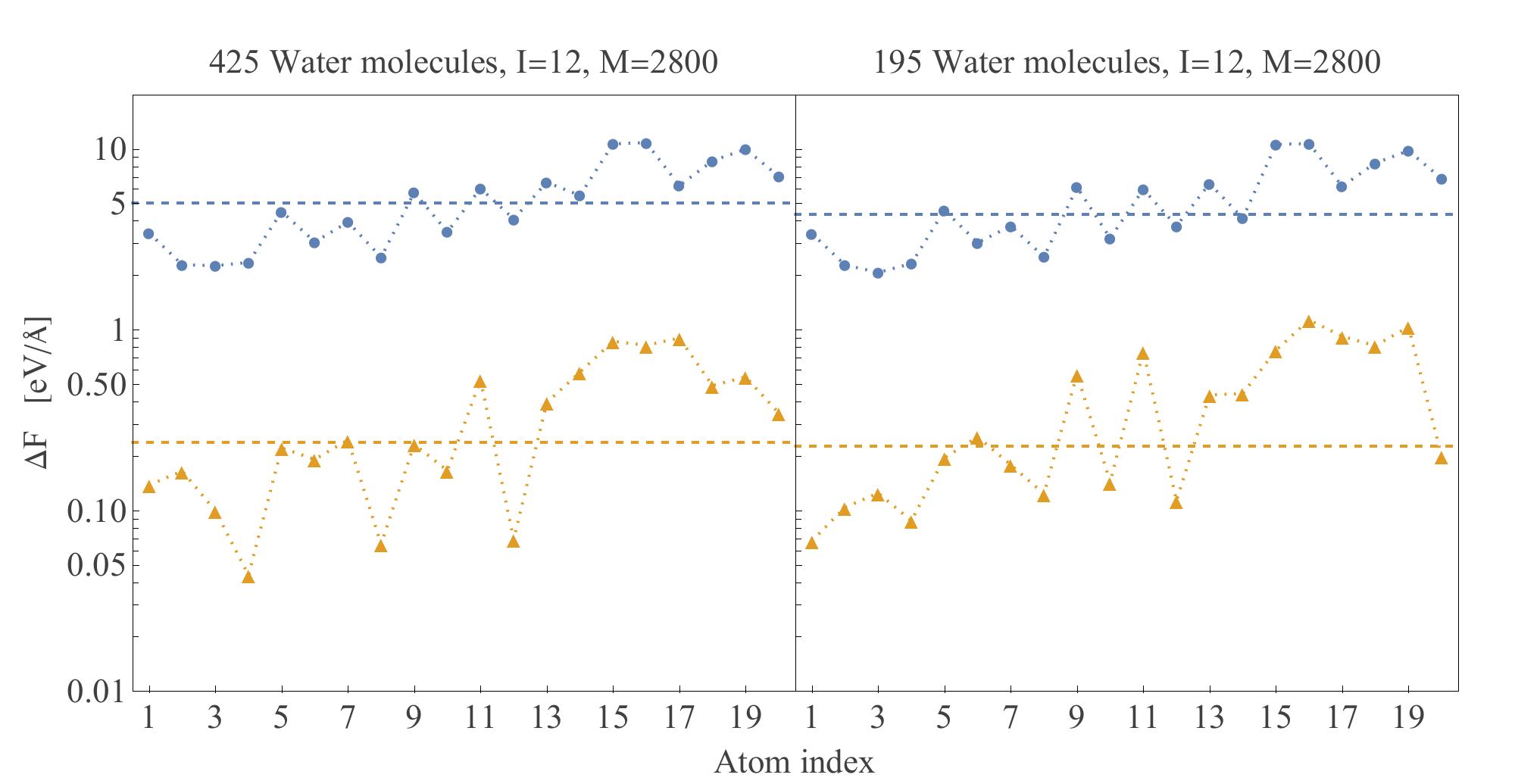}\caption{\label{fig:system comp}The statistical errors in the sDFT forces
acting on the $20$ nitrogen atoms in two solvated-TrpZip2 systems
(using 425 solvating water molecules on the left and 195 on the right).
Forces for both systems were calculated using $I=12$ stochastic vectors.
For each Nitrogen atom, we show the uncertainty $\sigma_{C}$ (blue
dots), and the estimate in the bias $\Delta\rho_{C}$ (orange triangles),
see Eqs. $27-28$ in manuscript. The dotted lines connecting the markers
are presented as a guide for the eye, while the dashed horizontal
lines are medians over all atoms of $\sigma_{C}$ and $\Delta\rho_{C}$.}
\end{figure}

\section{System size dependency}

We compare the statistical errors of sDFT forces of two Trp-zip2 peptide
systems, each with a different number of solvating water molecules.
The first is the system we presented results for in Section III of
the manuscript, solvated by $425$ water molecules, while the second
is a smaller system, with only $195$ solvating water molecules. To
keep the systems as similar as possible we cropped the smaller system
directly from the larger one, such that the $195$ solvating water
molecules that are closest to the peptide are identical in their geometry
in both systems, and the embedded-fragment method was employed such
that, for both systems, the entire peptide composes one fragment,
while the remaining water molecules are split into fragments with
an average number of $16$ water molecules per fragment.

In Fig. \ref{fig:system comp} we present the uncertainty and bias
estimates for the sDFT forces on the 20 nitrogen atoms of the peptide,
in the two systems described above, for the case of $I=12$ stochastic
orbitals. To allow for a comparison of the bias estimates, we repeated
the calculation a large, $M=2800$, times. The uncertainty values
of each nitrogen atom are near identical between the two systems.
Since the errors fluctuate, even for $2800$ repetitions, we compare
the median value over all nitrogen atoms, and find that it is practically
identical for the two systems. These results give indication to a
weak dependency of the statistical errors, uncertainty and bias, on
system size.

\section{Comparison with real-space grid representation }

We have checked the speed of basis sets vs. real-space grid calculations
(as described in Ref.~\citep{arnon2017equilibrium}) with $\text{Si}_{35}\text{H}_{36}$
as a benchmark. For the real-space grid calculations we used a grid
of $64^{3}$ points of spacing $\delta x=0.5a_{0}$ and the wall time
for the real-space sDFT/SCF iteration was one minute. For the basis-set
sDFT calculation we used the same integration grid. 

Here are the results of our analysis (summarized in Table~\ref{tab:Data-for-comparing}):
the Hamiltonian operation for STO-3G and 6-31G are a factor 40 and
4 respectively faster than that of the real-space grid. Furthermore,
the energy range of the basis sets is much smaller, so the Chebyshev
expansion length of STO-3G and 6-31G are a factor 13 and 8 shorter
than that of the real space grid. As a result, the operation of the
DM on a vector in STO-3G and 6-31G are a factor of about 500 and 30
faster respectively than in the real-space grid calculation. Therefore,
the calculation of a single SCF iteration is much faster in the basis
set calculations than in the real-space grid.

Next, we considered the fluctuation error in both calculations. Surprisingly,
we found, that the standard deviation of a typical force component
on Si, for the STO-3G and the 6-31G basis sets are larger by a factor
$1.4$ and $4.6$ respectively than that for the real-space grid.
This means that for a given standard deviation goal, the basis set
calculations will require more CPUs by factors of about $N_{p}=1.4^{2}\approx2$
and $N_{p}=4.6^{2}\approx21$, respectively, than the grid calculation. 

\begin{table}
\begin{tabular}{|c|>{\centering}m{0.1\textwidth}|>{\centering}m{0.1\textwidth}|>{\centering}m{0.1\textwidth}|>{\centering}m{0.1\textwidth}|>{\centering}m{0.12\textwidth}|>{\centering}m{0.14\textwidth}|>{\centering}m{0.14\textwidth}|}
\hline 
 & Relative speed of Hamiltonian operation & Relative Chebyshev expansion length & Relative speed & Relative statistical error in force & Relative required number of CPUs & \multicolumn{2}{c|}{Relative wall-time}\tabularnewline
\hline 
 & $v_{H}$ & $N_{C}$ & $v=\frac{v_{H}}{N_{C}}$ & $\delta f$ & $N_{p}=\delta f^{2}$ & Running on the same number of processors $N_{p}/v$ & Running on $N_{p}$ times more processors $1/v$\tabularnewline
\hline 
\hline 
STO-3G & 40 & 0.08 & 500 & 1.4 & 2 & 0.004 & $2\times10^{-3}$\tabularnewline
\hline 
6-31G & 4 & 0.13 & 31 & 4.6 & 21 & 0.700 & $3\times10^{-2}$\tabularnewline
\hline 
Real space & 1 & 1 & 1 & 1 & 1 & 1 & 1\tabularnewline
\hline 
\end{tabular}

\caption{\label{tab:Data-for-comparing}Data for comparing speeds and wall
times for real-space and basis-set calculations}
\end{table}

Summarizing, STO-3G calculations can be very fast relative to the
real-space grid's, however, it is well known that STO-3G is not accurate
enough to be a reliably useful basis. When run with the same number
of processors, the overall numerical effort of the 6-31G basis is
comparable to that of the real-space grid calculation. However, if
one can allocate a factor of $N_{p}\approx21$ more processors to
the 6-31G calculation than needed by the real-space grid, the overall
speed of the former will be 30 times faster. 

\bibliographystyle{aapmrev4-2}
\bibliography{forcesBib}